\documentclass[11pt,a4paper]{article}

\usepackage{amsfonts,amsmath,amssymb,amsthm}
\usepackage{latexsym,amscd}
\usepackage{amsbsy}
\usepackage{CJK}
\usepackage{indentfirst,mathrsfs}
\usepackage{latexsym,amscd}
\usepackage{amsbsy}
\usepackage[noadjust]{cite}
\usepackage{color}
\usepackage{multirow}
\usepackage{makecell}
\usepackage{float}
\usepackage{booktabs}
\usepackage{multirow}
\usepackage{authblk}
\usepackage{hyperref}

\newtheorem{theorem}{Theorem}
\newtheorem{lem}{Lemma}

\newtheorem{example}{Example}
\newtheorem{remark}{Remark}

\newcommand{\fp}{{\mathbb F}_{p}}

\newcommand{\Tr}{{\rm {Tr}}}

\newcommand{\C}{{\mathcal{C}}}

\begin{document}

\title{Several classes of few-weight linear codes based on Weil sums}

\author{Lan Deng
\thanks{L. Deng is with the Hubei Key Laboratory of Applied Mathematics, Faculty of Mathematics and Statistics, Hubei University, Wuhan, 430062, China, and also with the School of Mathematics and Statistics, Hanjiang Normal University, Shiyan, 442000, China. Email: denglan2026@126.com
},
Zhao Hu,
Nian Li
\thanks{Z. Hu and N. Li are with the Key Laboratory of Intelligent Sensing System and Security (Hubei University), Ministry of Education, the Hubei Provincial Engineering Research Center of Intelligent Connected Vehicle Network Security, School of Cyber Science and Technology, Hubei University, Wuhan 430062. Email: zhao.hu@aliyun.com, nian.li@hubu.edu.cn}
}

\date{}
\maketitle
\vspace{-4mm}
\begin{abstract}
 Linear codes over finite fields with few weights are of significant interest due to their applications in areas such as secret sharing schemes, strongly regular graphs, and association schemes. In this paper, we construct several classes of $p$-ary linear codes based on Weil sums via the defining-set construction of bivariate form, where $p$ is an odd prime. Utilizing the theory of quadratic forms and Weil sums over finite fields, the parameters and weight distributions of the codes are completely determined. As a result, several families of three-weight or four-weight linear codes are obtained from our construction. In addition, it is shown that most of our codes are minimal.
\end{abstract}

\section{Introduction}
Let $\mathbb{F}_q$ be the finite field with $q$ elements, and let $\mathbb{F}_q^* = \mathbb{F}_q \setminus \{0\}$, where $q = p^m$, $p$ is an odd prime and $m\geq 2$. An $[n,k,d]$ linear code $\C$ over $\mathbb{F}_q$ is a $k$-dimensional subspace of $\mathbb{F}_q^n$ with length $n$ and minimum distance $d$. For two codewords $\mathbf{a} = (a_1, a_2, \ldots, a_n)$ and $\mathbf{b} = (b_1, b_2, \ldots, b_n)$ in $\C$, the Hamming weight of a codeword $\mathbf{a}$ is $wt (\mathbf{a}) = |\{i : a_i \neq 0, 1 \leq i \leq n\}|$, and the Hamming distance between them is $d(\mathbf{a},\mathbf{b}) = wt (\mathbf{a}-\mathbf{b})$. The weight enumerator of a code $\mathcal{C}$ of length $n$ is the polynomial $W(z) = 1 + A_1 z + A_2 z^2 + \cdots + A_n z^n$, where $A_i$ denotes the number of codewords of Hamming weight $i$ in $\mathcal{C}$. The coefficient sequence $(1, A_1, \dots, A_n)$ is known as the weight distribution of the code. This distribution is of fundamental importance in coding theory, as it encodes essential information for evaluating the error-correcting capability of a code, as well as the probability of error detection and correction under various decoding algorithms \cite{Torl}. If exactly $t$ of the coefficients $A_1, A_2, \dots, A_n$ are non-zero, the code $\mathcal{C}$ is called a $t$-weight code. Linear codes with few weights can be used in secret sharing schemes \cite{ADHK,CDY,YD}, association schemes \cite{CG}, strongly regular graphs \cite{CK}, and authentication codes \cite{DHKW,DW}. Given the broad range of their applications, the systematic construction of few-weight linear codes and the precise determination of their weight distributions have become an active research direction in coding theory.

A general framework for constructing linear codes over finite fields was introduced by Ding and Niederreiter in 2007 \cite{DH}, relying on the use of trace functions. Let $D = \{d_1, d_2, \ldots, d_n\} \subseteq \mathbb{F}_q$ and let $\mathrm{Tr}$ be the trace function from $\mathbb{F}_q$ to $\mathbb{F}_p$. A linear code $\C_D$ of length $n$ over $\mathbb{F}_p$ is defined as
\[
\C_D = \{(\mathrm{Tr}(xd_1), \mathrm{Tr}(xd_2), \ldots, \mathrm{Tr}(xd_n)) : x \in \mathbb{F}_q\},
\]
where $D = \{d_1, d_2, \ldots, d_n\}$. The set $D$ is referred to as the defining set of the linear code $\C_D$, and accordingly this construction approach is called defining-set construction. This approach provides a flexible and unified way to produce linear codes with desirable properties. The choice of the defining set $D$ represents one of the central challenges in this area and has stimulated a great deal of research over the past decade. Many efforts have been devoted to constructing high-quality linear codes via the defining-set method, and various families of linear codes with interesting weight distributions have been obtained \cite{AKL,DC,DD,HLZWT,Duan,HQLT,MQRT,JLF,KK,MQCY,MOS,WLDL,ZQ,YKT}.

Later, the defining-set construction was extended to the bivariate form, namely, linear code $\C_{D}$ of the form
\begin{equation} \label{CD-1}
\C_D = \{\mathbf{c}(a,b) = (\mathrm{Tr}(ax + by))_{(x,y) \in D} : a, b \in \mathbb{F}_{q}\},
\end{equation}
where $D\subseteq \mathbb{F}_{q}\times \mathbb{F}_{q}$. Here $\mathbb{F}_{q}\times \mathbb{F}_{q}:=\{(x,y): x,y \in \mathbb{F}_{q}\}$.
In 2017, Li et al. \cite{LYF} presented several classes of two-weight and three-weight linear codes of the form \eqref{CD-1} by choosing the defining set
\[
D = \{(x,y) \in \mathbb{F}_q^2 \setminus \{(0,0)\} : \mathrm{Tr}(x^{l_1} + y^{l_2}) = 0\},
\]
where $l_1, l_2 \in \{ 1, 2, p^{\frac{m}{2} + 1} \}$.  In the same year, Zhu et al. \cite{ZX} constructed a new class of at most three-weight linear codes of the form \eqref{CD-1} by choosing the defining set
\[
D = \{(x,y) \in \mathbb{F}_{q}^2 \setminus \{(0,0)\} : \mathrm{Tr}(x^{p^{k_1}+1} + y^{p^{k_2}+1}) = c\},
\]
where $m/\gcd(m,k_1)$ is odd, $m/\gcd(m,k_2)$ is even, and $c \in \mathbb{F}_p$.
Subsequently, in 2021, Zheng et al. \cite{ZZWZ} further studied a class of linear codes $\C_{D}$ proposed by Zhu et al. \cite{ZX}, and determined their weight enumerators for any positive integers $m$, $k_1$ and $k_2$ .

In this paper, inspired by the previous works, we focus on the linear codes $\C_D$ of the form \eqref{CD-1} with the defining set
\begin{equation} \label{D-2}
D = \{(x, y) \in \mathbb{F}_{q}^2 \setminus \{(0,0)\} : \mathrm{Tr}(x^{p^{k_1}+1} + y^{p^{k_2}+1} + x + y) = c\},
\end{equation}
where $k_1$ and $k_2$ are non-negative integers and $c\in \mathbb{F}_p$. By using the quadratic forms and Weil sums over finite fields, the parameters and weight distributions of $\C_D$ are completely determined. Moreover, several classes of three-weight or four-weight linear codes $\C_D$ are derived. In addition, it is shown that most of our codes are minimal.

The remainder of this paper is organized as follows. In Section \ref{s2}, we introduce the necessary preliminaries, including fundamental concepts and auxiliary results on Gauss and Weil sums. Section \ref{s3} presents our main results. By considering three cases, we construct several classes of linear codes $\C_D$ over $\mathbb{F}_p$, and completely determine their parameters and weight distributions. The proofs of the main results are provided in Section \ref{s4}. In Section \ref{s5}, we investigate the minimality of these codes. Finally, the paper concludes with a summary in Section \ref{s6}.

\section{Preliminaries}\label{s2}
In this section, we briefly recall some basic results on finite fields that will be used throughout the paper. Unless otherwise stated, we adopt the following notation:
\begin{itemize}
\item Let $\mathbb{F}_{q}$ be the finite field with $q$ elements, where $q=p^m$, $p$ is an odd prime and $m\ge 2$.
\item Let $\mathrm{\rm SQ} = \{ x^2 \mid x \in \mathbb{F}_p^* \}$ and $\mathrm{\rm NSQ} = \mathbb{F}_p^* \setminus \mathrm{SQ}$.
\item Let $m_p \in \fp$ such that $m_p = m \mod p$.
\item Let $u = \gcd(m,k_1)$ and $v = \gcd(m,k_2)$, where $k_1$ and $k_2$ are non-negative integers.
\item Let $\xi_p = e^{\frac{2\pi\sqrt{-1}}{p}}$ be the primitive $p$-th root of unity.
\item Let $\mathrm{Tr}(\cdot)$ be the trace function from $\mathbb{F}_{q}$ to $\mathbb{F}_p$.
\item Let $v_2(\cdot)$ be the 2-adic order function and denote $v_2(0) = \infty$.
\item Let $\chi$ and $\chi_1$ be the canonical additive characters of $\mathbb{F}_{q}$ and $\mathbb{F}_p$, respectively.
\item Let $\eta$ and $\mathbb{\eta}_{1}$ be the quadratic multiplicative characters of $\mathbb{F}_q^*$ and $\mathbb{F}_p^*$, respectively. We extend these quadratic characters by letting $\eta(0)=0$ and $\mathbb{\eta}_{1}(0)=0$.
\end{itemize}

The Gauss sum over $\mathbb{F}_q$ is defined by

\[
G(\eta) = \sum_{x \in \mathbb{F}_{q}^*} \eta(x) \chi(x) = \sum_{x \in \mathbb{F}_{q}} \eta(x) \chi(x).
\]
Particularly, $G_1(\eta_1)$ denotes the Gauss sum over $\mathbb{F}_p$.

We now present several lemmas that play a crucial role in the proofs of our main results.

\begin{lem}{\rm\cite[Theorem 5.15]{LN}}\label{lemma1}
We have
\[
G(\eta) = (-1)^{m-1} (\sqrt{p^*} )^m
=
\begin{cases}
(-1)^{m-1} \sqrt{q}, & \mbox{ if } p \equiv 1 \bmod 4; \\
(-1)^{m-1} (\sqrt{-1})^m \sqrt{q}, & \mbox{ if } p \equiv 3 \bmod 4,
\end{cases}
\]
where $p^* = (-1)^{\frac{p-1}{2}} p$. In particular, $G_1(\eta_1)=\sqrt{p^*}$.
\end{lem}

In what follows, we recall a known general result on the exponential sums of quadratic functions from $\mathbb{F}_q$ to $\mathbb{F}_p$.

\begin{lem}{\rm\cite[Theorems 5.15 and 5.33]{LN}}\label{lemma2}
Let $R(x)$ be a quadratic function from $\mathbb{F}_{q}$ to $\mathbb{F}_p$ with rank $r \neq 0$. Then
\[
\sum_{x \in \mathbb{F}_{q}} \xi_p^{R(x)} =
\begin{cases}
\eta_1(\delta) p^{m - \frac{r}{2}}, & \text{if } p \equiv 1 \mod 4; \\
(-1)^{\frac{r}{2}} \eta_1(\delta) p^{m - \frac{r}{2}}, & \text{if } p \equiv 3 \mod 4,
\end{cases}
\]
where $\delta$ is the determinant of $R(x)$. Furthermore, for any $z \in \mathbb{F}_p^*$,
\[
\sum_{x \in \mathbb{F}_{q}} \xi_p^{zR(x)} = \eta_1^r(z) \sum_{x \in \mathbb{F}_{q}} \xi_p^{R(x)}.
\]
\end{lem}

\begin{lem}{\rm\cite[Theorem 5.33]{LN}}\label{lemma3}
Let $f(x) = a_2 x^2 + a_1 x + a_0 \in \mathbb{F}_{q}[x]$, where $a_2 \neq 0$. Then
\[
\sum_{x \in \mathbb{F}_{q}} \chi(f(x)) = \chi(a_0 - a_1^2 (4a_2)^{-1}) \eta(a_2) G(\eta, \chi).
\]
\end{lem}

\begin{lem}{\rm\cite{DHOU}}\label{lemma4}
Define $\mathrm{Tr}(x^{p^k+1})$ as the quadratic function from $\mathbb{F}_{q}$ to $\mathbb{F}_p$ that has a rank of $r$. Then
\[
r =
\begin{cases}
m - 2u, & \text{if } v_2(m) > v_2(k) + 1; \\
m, & \text{otherwise},
\end{cases}
\]
where $u = \gcd(m, k)$.
\end{lem}

We next consider Weil sums. These exponential sums provide the necessary tools for evaluating the character sums that emerge from the defining-set construction, and they are therefore fundamental to our work. For $a \in \mathbb{F}_{q}^*$ and $b \in \mathbb{F}_{q}$, define the Weil sum $S_k(a, b)$ by
\begin{equation} \label{S-3}
S_k(a, b) = \sum_{x \in \mathbb{F}_{q}} \chi( ax^{p^k+1} + bx ).
\end{equation}
In particular, when $b = 0$, the Weil sum $S_k(a, 0)$ is given by the following lemma.
\begin{lem}{\rm\cite[Corollary 7.6]{DHOU}}\label{lemma5}
For $a \in \mathbb{F}^*_{q}$, we have the following results.
\begin{itemize}
\item [1)] If $v_2(m) \leq v_2(k)$, then
\[
S_k(a,0) = \eta(a)(-1)^{m-1}(\sqrt{-1})^{\frac{(p-1)^2 m}{4}} p^{\frac{m}{2}}.
\]
\item [2)] If $v_2(m) = v_2(k) + 1$, then
\[
S_k(a,0) =
\begin{cases}
p^{\frac{m+\gcd(2k,m)}{2}}, & \text{if } a^{\frac{(p^k-1)(p^m-1)}{p^{\gcd(2k,m)}-1}} = -1; \\
-p^{\frac{m}{2}}, & \text{otherwise}.
\end{cases}
\]
\item [3)] If $v_2(m) > v_2(k) + 1$, then
\[
S_k(a,0) =
\begin{cases}
-p^{\frac{m+\gcd(2k,m)}{2}}, & \text{if } a^{\frac{(p^k-1)(p^m-1)}{p^{\gcd(2k,m)}-1}} = 1; \\
p^{\frac{m}{2}}, & \text{otherwise}.
\end{cases}
\]
\end{itemize}
\end{lem}

However, the analysis for $b \neq 0$ is more delicate. When $b \neq 0$, the value of $S_k(a, b)$ is determined by the solutions of the polynomial $a^{p^k} x^{p^{2k}} + ax = 0$. We begin by recalling the known result concerning this equation.

\begin{lem}{\rm\cite[Theorem 4.1]{CSR}}\label{lemma6}
Let $\alpha = \gcd(m,k)$ and $a \in \mathbb{F}_{q}^*$. Then the equation
\[
a^{p^k} x^{p^{2k}} + ax = 0
\]
is solvable in $\mathbb{F}_{q}^*$ if and only if $m / \alpha$ is even and $a^{\frac{p^{m}-1}{p^{\alpha}+1}} = (-1)^\frac{m}{2\alpha}$. In such cases, there are $p^{2\alpha} - 1$ non-zero solutions.
\end{lem}

Based on the above discussion, these two lemmas together determine the possible values of the Weil sum $S_k(a, b)$ under various conditions.
\begin{lem}{\rm\cite[Theorem 1]{Coul}}\label{lemma7}
Let $\alpha = \gcd(m,k)$. Let $f(x) = a^{p^k} x^{p^{2k}} + ax$ be a permutation polynomial over $\mathbb{F}_{q}$, where $a \in \mathbb{F}_{q}^*$. Assume that $x_0$ is the unique solution of the equation $f(x) = -b^{p^k}$. The following statements hold.
\begin{itemize}
\item [1)]  If $m/\alpha$ is odd, then
\[
S_k(a,b) = (-1)^{m-1} ( \sqrt{-1} )^{\frac{(p-1)^2}{4}3m} p^{\frac{m}{2}} \eta(a) \chi (-a x_0^{p^k+1})
\]
\[
=
\begin{cases}
(-1)^{m-1} p^{\frac{m}{2}} \eta(a) \chi (-a x_0^{p^k+1}), & \text{if } p \equiv 1 \mod 4; \\
(-1)^{m-1} ( \sqrt{-1} )^{3m} p^{\frac{m}{2}} \eta(a) \chi (-a x_0^{p^k+1}), & \text{if } p \equiv 3 \mod 4.
\end{cases}
\]

\item [2)]  If $m/\alpha$ is even, then
\[
a^{\frac{p^m-1}{p^\alpha+1}} \neq (-1)^{\frac{m}{2\alpha}}
\]
and
\[
S_k(a,b) = (-1)^{\frac{m}{2\alpha}} p^{\frac{m}{2}} \chi (-a x_0^{p^k+1}).
\]
\end{itemize}
\end{lem}
\begin{lem}{\rm\cite[Theorem 2]{Coul}}\label{lemma8}
Let $\alpha = \gcd(m,k)$ and let $m$ be even. Assume that $f(x) = a^{p^k} x^{p^{2k}} + ax$ is not a permutation polynomial over $\mathbb{F}_{q}$. Then $S_k(a,b) = 0$ unless the equation $f(x) = -b^{p^k}$ is solvable. If the equation has a solution $x_0$, then
\[
S_k(a,b) = -(-1)^{\frac{m}{2\alpha}} p^{\frac{m}{2}+\alpha} \chi (-a x_0^{p^k+1}).
\]
\end{lem}
In the following computations, the subsequent results will also be required.
\begin{lem}{\rm\cite[Lemma 13]{JLF}}\label{lemma9}
 Let $\alpha = \gcd(m,k)$. If $\frac{m}{\alpha} \equiv 0 \pmod{4}$, then
\[
| \{ z \in \mathbb{F}_{q} : x^{p^{2k}} + x = z^{p^k} \text{ is solvable in } \mathbb{F}_{q} \} | = p^{m-2\alpha}.
\]
\end{lem}

Next, we recall the Pless power moment identities of linear codes over $\fp$. They relate the weight enumerator of a linear code to that of its dual, and these identities often help determine the weight distributions of codes with only a few nonzero weights. Let $\C$ be an $[n, k]$ linear code over $ \mathbb{F}_p $ and denote its dual by $\C^{\perp}$. Let $A_i$ and $ A_i^{\perp} $ be the number of codewords of weight $ i $ in $ \mathcal{C} $ and $ \mathcal{C}^{\perp} $, respectively. The first two Pless power moment identities \cite[p.~256]{Huff} are given by
$\sum_{i=0}^{n} A_i = p^k$ and $\sum_{i=0}^{n} iA_i = p^{k-1}\big(pn - n - A_1^{\perp}\big)$.

Aschikhmin and Barg’s result, which is often used to tell whether a linear code over $\fp$ is minimal, is presented as
follows.

\begin{lem}{\rm\cite{{Ashikhmin}}}\label{lemma11}
A linear code $\C$ over $\fp$ is minimal if $\frac{w_{\min}}{w_{\max}} > \frac{p-1}{p}$, where $w_{\min}$ and $w_{\max}$ denote the minimum and maximum nonzero weights in $\C$, respectively.
\end{lem}

\section{Main results} \label{s3}
In this section, we present our main results. Note that the proofs of these results will be given in Section \ref{s4}.

We first recall the defining set in \eqref{D-2} given by
\[
D = \{(x, y) \in \mathbb{F}_q^2 \setminus \{(0,0)\}: \mathrm{Tr}(x^{p^{k_1}+1} + y^{p^{k_2}+1} + x + y) = c\},
\]
where $p$ is an odd prime and $c \in \mathbb{F}_p$.

In what follows, we investigate the linear code $\C_D$ defined as in \eqref{CD-1} and \eqref{D-2} by considering the three cases: 1) $2v_2(m) = v_2(u) + v_2(v)$; 2) $2v_2(m) = v_2(u) + v_2(v) + 1$; 3) $2v_2(m) > v_2(u) + v_2(v) + 1$, where $u = \gcd(m, k_1)$ and $v = \gcd(m, k_2)$. For convenience, we define the following notation:
\begin{equation} \label{v1}
\varepsilon_u =
\begin{cases}
1, & \text{if } v_2(m) > v_2(u) + 1; \\
0, & \text{if } v_2(m) \leq v_2(u) + 1,
\end{cases}
\quad\varepsilon_v =
\begin{cases}
1, & \text{if } v_2(m) > v_2(v) + 1; \\
0, & \text{if } v_2(m) \leq v_2(v) + 1,
\end{cases}
\end{equation}
and
\begin{equation} \label{v3}
\delta_c =
\begin{cases}
1, & \text{if } c=0; \\
0, & \text{if } c \in \mathbb{F}_p^*,
\end{cases}
\quad\sigma_c =
\begin{cases}
1, & \text{if } \frac{m_p}{2} + c \neq 0, {m_p} \neq 0 \text{ and } \eta_1 ( \frac{2c}{m_p + 2c} ) \neq -1; \\
0, & \text{otherwise}.
\end{cases}
\end{equation}
\begin{theorem}\label{thm1}
Let $\C_D$ be defined by \eqref{CD-1} and \eqref{D-2}, $\delta_c$ and $\sigma_c$ be given as in \eqref{v3}, and $\tau_1=(-1)^{\frac{m(p-1)}{2}}p^{m-1}$.
Assume that $2v_2(m) = v_2(u) + v_2(v)$. \\
1) If $\frac{m_p}{2} + c = 0$, then $\C_D$ is a 3-weight $[ p^{2m-1} + \tau_1(p-1)-\delta_c, 2m ]$ linear code over $\mathbb{F}_p$ with the weight distribution
\begin{center}
\begin{tabular}{ll}
\toprule
Weight $w$ & Frequency $A_w$\\
\midrule
0 & 1 \\
$p^{2m-1} - p^{2m-2}$ & $p^{2m-2} +\delta_c(p-1)\tau_1 - 1$ \\
$p^{2m-1} - p^{2m-2} + (p-1)\tau_1$ & $(p-1)(2p^{2m-2} + (1-2\delta_c)\tau_1)$ \\
$p^{2m-1} - p^{2m-2} + (p-2)\tau_1$ & $ p^{2m-2} (p-1)^2 + (\delta_c-1)(p-1)\tau_1 $ \\
\bottomrule
\end{tabular}
\end{center}
2) If $\frac{m_p}{2} + c \neq 0$, then $\C_D$ is a 3-weight $[ p^{2m-1} - \tau_1-\delta_c, 2m ]$ linear code over $\mathbb{F}_p$ with the weight distribution
\begin{center}
\begin{tabular}{ll}
\toprule
Weight $w$ & Frequency $A_w$ \\
\midrule
0 & 1 \\
$p^{2m-1} - p^{2m-2}$ & $\frac{1}{2} (p^{2m-2}(p^2-p+2) +(p-1)\tau_1(1-2\sigma_c-\sigma_c\delta_c(p-1))) -1$ \\
$p^{2m-1} - p^{2m-2} - \tau_1$ & $(p-1)(2p^{2m-2} + \sigma_c(2-\delta_c)\tau_1)$ \\
$p^{2m-1} - p^{2m-2} - 2\tau_1$ & $\frac{1}{2}(p-1)(p^{2m-2}(p-2)+(\sigma_c\delta_c(p+1)-2\sigma_c-1)\tau_1) $ \\
\bottomrule
\end{tabular}
\end{center}
\end{theorem}


\begin{example}
Let $p = 5$, $m = 3$, $k_1 = 1$, $k_2 = 2$, and $c = 1$. Magma experiments show that $\C_D$ has parameters $[3225,6,2500]$ with weight enumerator $1 + 624x^{2500} + 9900x^{2575} + 5100x^{2600}$, which is consistent with Theorem \ref{thm1}-1).
\end{example}

\begin{example}
Let $p = 3$, $m = 5$, $k_1 = k_2 = 1$, and $c = 1$. Magma experiments show that $\C_D$ has parameters $[19764,10,13122]$ with weight enumerator $1 + 26244x^{13203} + 6642x^{13284} + 26162x^{13122}$, which is consistent with Theorem \ref{thm1}-2).
\end{example}




\begin{theorem}\label{thm22}
Let $\C_D$ be defined by \eqref{CD-1} and \eqref{D-2}, $\delta_c$ and $\sigma_c$ be given as in \eqref{v3}, and $\tau_2=(-1)^{\frac{m(p-1)}{4}}p^{m-1}$.
Assume that $2v_2(m) = v_2(u) + v_2(v)+1$. \\
1) If $\frac{m_p}{2} + c = 0$, then $\C_D$ is a 3-weight $[ p^{2m-1} + \tau_2(p-1)-\delta_c, 2m ]$ linear code over $\mathbb{F}_p$ with the weight distribution
\begin{center}
\begin{tabular}{ll}
\toprule
Weight $w$ & Frequency $A_w$\\
\midrule
0 & 1 \\
$p^{2m-1} - p^{2m-2}$ & $p^{2m-2} +\delta_c(p-1)\tau_2 - 1$ \\
$p^{2m-1} - p^{2m-2} + (p-1)\tau_2$ & $(p-1)(2p^{2m-2} + (1-2\delta_c)\tau_2)$ \\
$p^{2m-1} - p^{2m-2} + (p-2)\tau_2$ & $ p^{2m-2} (p-1)^2 + (\delta_c-1)(p-1)\tau_2 $ \\
\bottomrule
\end{tabular}
\end{center}
2) If $\frac{m_p}{2} + c \neq 0$, then $\C_D$ is a 3-weight $[ p^{2m-1} - \tau_2-\delta_c, 2m ]$ linear code over $\mathbb{F}_p$ with the weight distribution
\begin{center}
\begin{tabular}{ll}
\toprule
Weight $w$ & Frequency $A_w$ \\
\midrule
0 & 1 \\
$p^{2m-1} - p^{2m-2}$ & $\frac{1}{2} (p^{2m-2}(p^2-p+2) +(p-1)\tau_2(1-2\sigma_c-\sigma_c\delta_c(p-1))) -1$ \\
$p^{2m-1} - p^{2m-2} - \tau_2$ & $(p-1)(2p^{2m-2} + \sigma_c(2-\delta_c)\tau_2)$ \\
$p^{2m-1} - p^{2m-2} - 2\tau_2$ & $\frac{1}{2}(p-1)(p^{2m-2}(p-2)+(\sigma_c\delta_c(p+1)-2\sigma_c-1)\tau_2) $ \\
\bottomrule
\end{tabular}
\end{center}
\end{theorem}

\begin{example}
Let $p = 3$, $m = 2$, $k_1 =2$, $k_2 = 1$, and $c = 2$. Magma experiments show that $\C_D$ has parameters $[21,4,12]$ with the weight enumerator $1 + 30x^{12} + 42x^{15} + 8x^{18}$, which is consistent with Theorem \ref{thm22}-1). This code is optimal due to \cite{Grassl}.
\end{example}




\begin{example}
Let $p = 5$, $m = 4$, $k_1 = 0$, $k_2 = 2$, and $c = 4$. Magma experiments show that $\C_D$ has parameters $[78000,8,62250]$ with weight enumerator $1 + 171624x^{62500} + 126000x^{62375} + 93000x^{62250}$, which is consistent with Theorem \ref{thm22}-2).
\end{example}

\begin{theorem}\label{thm33}
Let $\C_D$ be defined by \eqref{CD-1} and \eqref{D-2}, $\delta_c$ and $\sigma_c$ be given as in \eqref{v3}, and $\tau_3=p^{m-(\varepsilon_u u+\varepsilon_v v) - 1}$. Assume that $2v_2(m) > v_2(u) + v_2(v) + 1$ and $m \geq \varepsilon_u u + \varepsilon_v v + 2$, where $\varepsilon_u $ and $\varepsilon_v$ are given as in \eqref{v1}. \\
1) If $\frac{m_p}{2} + c = 0$, then $\C_D$ is an at most 4-weight $[ p^{2m-1} + p^{m+\varepsilon_u u + \varepsilon_v v - 1}(p-1)-\delta_c, 2m, p^{2m-1} - p^{2m-2}]$ linear code over $\mathbb{F}_p$ with the weight distribution
\begin{center}
\begin{tabular}{ll}
\toprule
Weight $w$ & Frequency $A_w$ \\
\midrule
0 & 1 \\
$p^{2m-1} - p^{2m-2}$ & $\tau_3^2+\delta_c(p-1)\tau_3-1$ \\
$p^{2m-1} - p^{2m-2} + (p-1)p^{m+\varepsilon_u u + \varepsilon_v v - 1}$ & $(p-1)(2\tau_3^2 + (1-2\delta_c)\tau_3)$ \\
$p^{2m-1} - p^{2m-2} + (p-2)p^{m+\varepsilon_u u + \varepsilon_v v - 1}$ & $ \tau_3^2 (p-1)^2 + (\delta_c-1)(p-1)\tau_3 $ \\
$(p-1)(p^{2m-2} +p^{m+\varepsilon_u u + \varepsilon_v v - 1} - p^{m+\varepsilon_u u + \varepsilon_v v - 2})$ & $p^{2m} - p^{2m-2(\varepsilon_u u+\varepsilon_v v)}$ \\
\bottomrule
\end{tabular}
\end{center}
2) If $\frac{m_p}{2} + c  \neq 0$, then $\C_D$ is an at most 4-weight $[p^{2m-1} - p^{m+\varepsilon_u u+\varepsilon_v v - 1}-\delta_c, 2m,p^{2m-1} - p^{2m-2} - 2p^{m+\varepsilon_u u+\varepsilon_v v-1}]$ linear code over $\mathbb{F}_p$ with the weight distribution
\begin{center}
\begin{tabular}{ll}
\toprule
Weight $w$ & Frequency $A_w$ \\
\midrule
0 & 1 \\
$p^{2m-1} - p^{2m-2}$ & $\frac{1}{2}( \tau_3^2(p^2-p+2) +(p-1)\tau_3(1-2\sigma_c-\sigma_c\delta_c(p-1)))-1$ \\
$p^{2m-1} - p^{2m-2} - p^{m+\varepsilon_u u+\varepsilon_v v-1}$ & $(p-1)(2\tau_3^2 + \sigma_c(2-\delta_c)\tau_3)$ \\
$p^{2m-1} - p^{2m-2} - 2p^{m+\varepsilon_u u+\varepsilon_v v-1}$ & $\frac{1}{2}(p-1)(\tau_3^2(p-2)+(\sigma_c\delta_c(p+1)-2\sigma_c-1)\tau_3)$ \\
$(p-1)(p^{2m-2} -  p^{m+\varepsilon_u u+\varepsilon_v v-2})$ & $p^{2m} - p^{2m-2(\varepsilon_u u+\varepsilon_v v)}$ \\
\bottomrule
\end{tabular}
\end{center}
\end{theorem}


\begin{example}
Let $p = 3$, $m = 4$, $c = 1$, $k_1 = 3$, and $k_2 = 1$, which implies $\varepsilon_u = \varepsilon_v = 1$. Magma experiments show that $\C_D$ has parameters $[2673,8,1458]$ with weight enumerator $1 + 6480x^{1782} + 30x^{1701} + 8x^{1458} + 42x^{1944}$, which is consistent with Theorem \ref{thm33}-1).
\end{example}



\begin{example}
Let $p = 5$, $m = 4$, $c = 0$, and $k_1 = k_2 = 1$, which implies $\varepsilon_u = \varepsilon_v = 1$. Magma experiments show that $\C_D$ has parameters $[74999,8,56250]$ with weight enumerator $1 + 224x^{62500} + 220x^{59375} + 180x^{56250} + 390000x^{60000}$, which is consistent with Theorem \ref{thm33}-2).
\end{example}


\begin{remark}
When $\varepsilon_u = \varepsilon_v = 0$, the linear codes $\C_D$ in Theorems \ref{thm33}-1) and  \ref{thm33}-2) are reduced to $3$-weight linear codes.
\end{remark}

\begin{remark}
Compared with the code constructions in \cite{ZX} and \cite{ZZWZ}, the codes constructed in this paper, as defined in \eqref{CD-1} and \eqref{D-2}, include two additional linear terms in $x$ and $y$ in the defining set $D$.
Note that the codes in \cite{ZX} and \cite{ZZWZ} are linear codes with two or three weights, whose parameters and weight distributions are independent of $c$. In contrast, our codes have three or four weights, and both their parameters and weight distributions depend on $c$.
Moreover, \cite{ZX} considers only the case where $m/\gcd(m,k_1)$ is odd and $m/\gcd(m,k_2)$ is even, whereas this paper investigates the general case without any restrictions.
\end{remark}

\begin{remark}
Determining whether two codes are equivalent is generally difficult. Equivalent codes necessarily have the same parameters and weight distribution, though the reverse is not always true. We therefore compare our codes with previously known ones only by their parameters and weight distributions. Some linear codes with $t$-weight were presented in \cite{CG,ChengC,DD,HQLT,JLF,LWL,LYF,LDQ,LIUDX,Sinak,MOS,TXF,WLDL,WLZ,ZZWZ}, where $ t \leq 4$. By comparing our codes with those in the above references, we find that our codes differ from these codes. For some special cases, our codes in Theorems \ref{thm1} and \ref{thm22} can be reduced to the codes in some of these references, which are shown as follows:
\begin{itemize}
\item [1)] When $c=0$ and $(-1)^{\frac{m(p-1)}{2}}=1$ (resp. $(-1)^{\frac{m(p-1)}{4}}=1$), our codes in Theorem \ref{thm1}-1) (resp. Theorem \ref{thm22}-1)) have the same parameters and weight distributions as those in \cite[Theorem 1-(1)]{LDQ} and \cite[Theorem 1-(1)] {Duan};
\item [2)] When $c \in \mathbb{F}_p^*$ and $(-1)^{\frac{m(p-1)}{2}}=1$ (resp. $(-1)^{\frac{m(p-1)}{4}}=1$), our codes in Theorem \ref{thm1}-1) (resp. Theorem \ref{thm22}-1)) have the same parameters and weight distributions as those in \cite[Theorem 1-(2)] {LDQ};
\item [3)] When $c \in \mathbb{F}_p^*$, $\sigma_c=0$, and $(-1)^{\frac{m(p-1)}{2}}=1$ (resp. $(-1)^{\frac{m(p-1)}{4}}=1$), our codes in Theorem \ref{thm1}-2) (resp. Theorem \ref{thm22}-2)) have the same parameters and weight distributions as those in \cite[Theorem 1-(2)] {LIUDX}.
\end{itemize}
\end{remark}

\section{Proofs of main results} \label{s4}
In this section, we present the proofs of our main results.  We first provide some auxiliary results.

\subsection{Some auxiliary lemmas}

In this subsection, we give some lemmas to compute the parameters and weight distributions of the linear codes defined by \eqref{CD-1} and \eqref{D-2}.

The length of the code  $\C_D$ relies on the following lemma.

\begin{lem}\label{lemma12}
Let $\C_D$ be a linear code defined by \eqref{CD-1} and \eqref{D-2}. Let $n:= |D|$ be the length of $\C_D$ and $\delta_c$ be given as in \eqref{v3}.\\
1) If $2v_2(m) = v_2(u) + v_2(v)$, then
  \begin{align}
  n&=\left\{\begin{array}{ll}
p^{2m-1} + (-1)^{\frac{m(p-1)}{2}} p^{m-1}(p-1)-\delta_c, & {\rm if} \,\, \frac{m_p}{2} + c = 0;\\
p^{2m-1} - (-1)^{\frac{m(p-1)}{2}} p^{m-1}-\delta_c, & {\rm if} \,\, \frac{m_p}{2} + c \neq 0.
\end{array}\right.\nonumber
  \end{align}
2) If $2v_2(m) = v_2(u) + v_2(v) + 1$, then
  \begin{align}
  n&=\left\{\begin{array}{ll}
p^{2m-1} + (-1)^{\frac{m(p-1)}{4}} p^{m-1}(p-1)-\delta_c, & {\rm if} \,\, \frac{m_p}{2} + c = 0;\\
p^{2m-1} - (-1)^{\frac{m(p-1)}{4}} p^{m-1}-\delta_c, & {\rm if} \,\, \frac{m_p}{2} + c \neq 0 .
\end{array}\right.\nonumber
  \end{align}
3) If $2v_2(m) > v_2(u) + v_2(v) + 1$, then
  \begin{align}
  n&=\left\{\begin{array}{ll}
p^{2m-1} + p^{m+\varepsilon_u u+\varepsilon_v v-1}(p-1)-\delta_c, & {\rm if} \,\, \frac{m_p}{2} + c = 0;\\
p^{2m-1} - p^{m+\varepsilon_u u+\varepsilon_v v-1}-\delta_c, & {\rm if} \,\, \frac{m_p}{2} + c \neq 0.
\end{array}\right.\nonumber
  \end{align}

\end{lem}

\begin{proof}
By the orthogonal property of additive characters, we have
\begin{align*}
n &= | \{ (x, y) \in \mathbb{F}_q^2 \setminus \{(0, 0)\}  : \Tr(x^{p^{k_1}+1} + y^{p^{k_2}+1} + x + y) = c \} | \\
&= \sum_{(x, y) \in \mathbb{F}_q^2 \setminus \{(0, 0)\}} \frac{1}{p} \sum_{z \in \mathbb{F}_p} \xi_p^{z ( \Tr(x^{p^{k_1}+1} + y^{p^{k_2}+1} + x + y) - c )} \\
&= \sum_{x,y \in \mathbb{F}_q} \frac{1}{p} ( 1 + \sum_{z \in \mathbb{F}_p^*} \xi_p^{z ( \Tr(x^{p^{k_1}+1} + y^{p^{k_2}+1} + x + y) - c )} )- \frac{1}{p} \sum_{z \in \mathbb{F}_p} \xi_p^{-cz} \\
&= p^{2m-1} + \frac{1}{p} \sum_{z \in \mathbb{F}_p^*} \xi_p^{-zc} \sum_{x \in \mathbb{F}_q} \chi( zx^{p^{k_1}+1} + zx ) \sum_{y \in \mathbb{F}_q} \chi( zy^{p^{k_2}+1} + zy )-\delta_c.
\end{align*}
For $z \in \mathbb{F}_p^*$, it can be verified that $-\frac{1}{2}$ is a solution of $z^{p^k} x^{p^{2k}} + zx = -z^{p^k}$.

1). Due to $2v_2(m) = v_2(u) + v_2(v)$, we have $v_2(m) = v_2(u) = v_2(v)$.
Since $\frac{m}{u}$ and $\frac{m}{v}$ are odd, one can verify that $h_{k_1}(x) = z^{p^{k_1}}x^{p^{2{k_1}}} + zx$ and $h_{k_2}(x) = z^{p^{k_2}}x^{p^{2{k_2}}} + zx$ are permutations over $\mathbb{F}_{q}$ for any $z \in \mathbb{F}_p^*$. Then, by Lemma \ref{lemma7}-1), it leads to
\begin{align*}
n &= p^{2m-1} + (-1)^{\frac{m(p-1)}{2}} p^{m-1} \sum_{z \in \mathbb{F}_p^*} \xi_p^{-z ( m_p ( -\frac{1}{2} )^{p^{k_1}+1} + m_p ( -\frac{1}{2} )^{p^{k_2}+1} + c ) }-\delta_c\\
&= p^{2m-1} + (-1)^{\frac{m(p-1)}{2}} p^{m-1} \sum_{z \in \mathbb{F}_p^*} \xi_p^{-z( \frac{m_p}{2} + c )}-\delta_c  \\
&=
\begin{cases}
p^{2m-1} + (-1)^{\frac{m(p-1)}{2}} p^{m-1}(p-1)-\delta_c, & \text{if } \frac{m_p}{2} + c = 0; \\
p^{2m-1} - (-1)^{\frac{m(p-1)}{2}} p^{m-1}-\delta_c, & \text{if } \frac{m_p}{2} + c \neq 0 .
\end{cases}
\end{align*}

2). Due to $2v_2(m) = v_2(u) + v_2(v) + 1$, $m$ is even, and thus there are two cases:
\begin{equation}\label{W2}
\left\{
\begin{aligned}
&v_2(m) = v_2(u) \text{ and } v_2(m) = v_2(v) + 1; \\
&v_2(m) = v_2(u)+1 \text{ and } v_2(m) = v_2(v).
\end{aligned}
\right.
\end{equation}

Next we only prove the case $v_2(m) = v_2(u)$ and $v_2(m) = v_2(v) + 1$, since the other case can be similarly proved. This implies that $\frac{m}{u}$ and $\frac{m}{2v}$ are odd. Clearly, for any $z \in \mathbb{F}_p^*$, $h_{k_1}(x) = z^{p^{k_1}}x^{p^{2{k_1}}} + zx$ is a permutation and by Lemma \ref{lemma6}, $h_{k_2}(x) = z^{p^{k_2}}x^{p^{2{k_2}}} + zx$ is also a permutation.
By Lemma \ref{lemma7}, it gives
\begin{align*}
n &= p^{2m-1} + (-1)^{\frac{m(p-1)}{4}} p^{m-1} \sum_{z \in \mathbb{F}_p^*} \xi_p^{-z ( m_p ( -\frac{1}{2} )^{p^{k_1}+1} + m_p ( -\frac{1}{2} )^{p^{k_2}+1} + c ) }-\delta_c\\
&= p^{2m-1} + (-1)^{\frac{m(p-1)}{4}} p^{m-1} \sum_{z \in \mathbb{F}_p^*} \xi_p^{-z( \frac{m_p}{2} + c )}-\delta_c  \\
&=
\begin{cases}
p^{2m-1} + (-1)^{\frac{m(p-1)}{4}} p^{m-1}(p-1)-\delta_c, & \text{if } \frac{m_p}{2} + c = 0; \\
p^{2m-1} - (-1)^{\frac{m(p-1)}{4}} p^{m-1}-\delta_c, & \text{if } \frac{m_p}{2} + c \neq 0.
\end{cases}
\end{align*}

3). Due to $2v_2(m) > v_2(u) + v_2(v) + 1$, $m$ is even and there are six cases:
\begin{equation}\label{W1}
\left\{
\begin{aligned}
&v_2(m) = v_2(u) \text{ and } v_2(m) > v_2(v) + 1; \\
&v_2(m) = v_2(v) \text{ and } v_2(m) > v_2(u) + 1; \\
&v_2(m) = v_2(u) + 1 \text{ and } v_2(m) = v_2(v) + 1; \\
&v_2(m) = v_2(u) + 1 \text{ and } v_2(m) > v_2(v) + 1; \\
&v_2(m) > v_2(u) + 1 \text{ and } v_2(m) = v_2(v) + 1; \\
&v_2(m) > v_2(u) + 1 \text{ and } v_2(m) > v_2(v) + 1.
\end{aligned}
\right.
\end{equation}

 Here we only prove the case $v_2(m) > v_2(u) + 1$ and $v_2(m) > v_2(v) + 1$, since the other cases can be similarly proved. This shows that $\varepsilon_u = 1$ and $\varepsilon_v = 1$. By Lemma \ref{lemma6}, $h_{k_1}(x) = z^{p^{k_1}}x^{p^{2{k_1}}} + zx$ and $h_{k_2}(x) = z^{p^{k_2}}x^{p^{2{k_2}}} + zx$ are not permutations over $\mathbb{F}_{q}$ for any $z \in \mathbb{F}_p^*$. By Lemma \ref{lemma8}, one has
\begin{align*}
n &= p^{2m-1} + p^{m+u+v-1} \sum_{z \in \mathbb{F}_p^*} \xi_p^{-z( \frac{m_p}{2} + c )}-\delta_c  \\
&=
\begin{cases}
p^{2m-1} + p^{m+u+v-1}(p-1)-\delta_c, & \text{if } \frac{m_p}{2} + c = 0; \\
p^{2m-1} - p^{m+u+v-1}-\delta_c, & \text{if } \frac{m_p}{2} + c \neq 0.
\end{cases}
\end{align*}

This completes the proof.
\end{proof}

The Hamming weight of the codeword $c(a, b) = ( \Tr(ax + by) )_{(x, y) \in D}$ in $\C_D$ is given by $wt (c(a, b))  = n - N(a, b)$, where $N(a, b)$ denotes the number of zero components in the codeword $c(a, b)$.
Then we have
\begin{align}\label{W4.2}
N(a, b) =& | \{ (x, y) \in D : \Tr(ax + by) = 0  \} |\nonumber\\
= & \sum_{(x, y) \in \mathbb{F}_q^2 \setminus \{(0, 0)\}}  ( \frac{1}{p} \sum_{z_1 \in \mathbb{F}_p} \xi_p^{z_1 (\Tr(x^{p^{k_1}+1}+y^{p^{k_2}+1}+x+y)-c)} ) ( \frac{1}{p} \sum_{z_2 \in \mathbb{F}_p} \xi_p^{z_2 \Tr(ax+by)} )\nonumber\\
= & \frac{1}{p^2} \sum_{x,y \in \mathbb{F}_q} (1 + \sum_{z_1 \in \mathbb{F}_p^*} \xi_p^{z_1 (\Tr(x^{p^{k_1}+1}+y^{p^{k_2}+1}+x+y)-c)} + \sum_{z_2 \in \mathbb{F}_p^*} \xi_p^{z_2 \Tr(ax+by)} \nonumber\\
&\quad + \sum_{z_1 \in \mathbb{F}_p^*} \xi_p^{z_1 (\Tr(x^{p^{k_1}+1}+y^{p^{k_2}+1}+x+y)-c)} \sum_{z_2 \in \mathbb{F}_p^*} \xi_p^{z_2 \Tr(ax+by)} ) - \frac{1}{p^2} \sum_{z_1,z_2 \in \mathbb{F}_p}\xi_p^{-z_1c} \nonumber\\
= &p^{2m-2} + \frac{1}{p^2} (\Omega_1 + \Omega_2 + \Omega_3)-\delta_c,
\end{align}
where
\begin{equation}\label{W4.3}
\Omega_1 = \sum_{x,y \in \mathbb{F}_q} \sum_{z_1 \in \mathbb{F}_p^*} \xi_p^{z_1 (\Tr(x^{p^{k_1}+1}+y^{p^{k_2}+1}+x+y)-c)},
\end{equation}
\[\Omega_2 = \sum_{x,y \in \mathbb{F}_q} \sum_{z_2 \in \mathbb{F}_p^*} \xi_p^{z_2 \Tr(ax + by)},\]
and
\[\Omega_3 = \sum_{x,y \in \mathbb{F}_q} \sum_{z_1 \in \mathbb{F}_p^*} \xi_p^{z_1(\Tr(x^{p^{k_1}+1} + y^{p^{k_2}+1} + x + y) - c)} \sum_{z_2 \in \mathbb{F}_p^*} \xi_p^{z_2 \Tr(ax + by)}. \]

From Lemma \ref{lemma12}, the value of $\Omega_1$ can be readily obtained. Moreover, it yields
\begin{align}
\Omega_2
&= \sum_{z_2 \in \mathbb{F}_p^*} \sum_{x \in \mathbb{F}_q} \xi_p^{\Tr(z_2 ax)} \sum_{y \in \mathbb{F}_q} \xi_p^{\Tr(z_2 by)} =
\begin{cases}
(p-1)q^2, & \text{if } (a,b) = (0,0); \\
0, & \text{if } (a,b) \neq (0,0).
\end{cases}
\label{W4.4}
\end{align}
Furthermore, a detailed computation gives
\begin{align*}
\Omega_3 &= \sum_{x,y \in \mathbb{F}_q} \sum_{z_1 \in \mathbb{F}_p^*} \xi_p^{z_1(\Tr(x^{p^{k_1}+1} + y^{p^{k_2}+1} + x + y) - c)} \sum_{z_2 \in \mathbb{F}_p^*} \xi_p^{z_2 \Tr(ax + by)} \\
&= \sum_{z_1, z_2 \in \mathbb{F}_p^*} \xi_p^{-z_1 c} \sum_{x \in \mathbb{F}_q} \xi_p^{\Tr(z_1 x^{p^{k_1}+1} + z_2 ax + z_1 x)} \sum_{y \in \mathbb{F}_q} \xi_p^{\Tr(z_1 y^{p^{k_2}+1} + z_2 by + z_1 y)} \\
&= \sum_{z_1, z_2 \in \mathbb{F}_p^*} \xi_p^{-z_1 c} \sum_{x \in \mathbb{F}_q} \xi_p^{\Tr(z_1 x^{p^{k_1}+1} + (z_2 a + z_1)x)} \sum_{y \in \mathbb{F}_q} \xi_p^{\Tr(z_1 y^{p^{k_2}+1} + (z_2 b + z_1)y)} \\
&= \sum_{z_1 \in \mathbb{F}_p^*} \xi_p^{-z_1 c} \sum_{z_2 \in \mathbb{F}_p^*} S_{k_1}(z_1, z_1 + z_2 a) S_{k_2}(z_1, z_1 + z_2 b),
\end{align*}
where $S_{k_1}(z_1, z_1 + z_2 a)$ and $S_{k_2}(z_1, z_1 + z_2 b)$ are Weil sums defined in \eqref{S-3}.

Next, we focus on the value of $\Omega_3$.

\begin{lem}\label{lemma13}
Let $(a,b)\in\mathbb F_q^2\setminus\{(0,0)\}$. Let $\gamma_a$ and $\gamma_b$ be solutions of the equations  $x^{p^{2{k_1}}} + x = -a^{p^{k_1}}$ and $x^{p^{2{k_2}}} + x = -b^{p^{k_2}}$, respectively. Assume that $2v_2(m) = v_2(u) + v_2(v)$.\\
1) If $\frac{m_p}{2} + c = 0$, then
\[
\Omega_3 =
\begin{cases}
(-1)^{\frac{m(p-1)}{2}} (p-1)^2 p^m, & \text{if } \Tr(\gamma_a^{p^{k_1}+1} + \gamma_b^{p^{k_2}+1}) = 0 ,\Tr(\gamma_a + \gamma_b) = 0; \\
-(-1)^{\frac{m(p-1)}{2}} (p-1)p^m, & \text{if } \Tr(\gamma_a^{p^{k_1}+1} + \gamma_b^{p^{k_2}+1}) = 0 , \Tr(\gamma_a + \gamma_b) \neq 0, \\
& \quad\text{or } \Tr(\gamma_a^{p^{k_1}+1} + \gamma_b^{p^{k_2}+1}) \neq 0 ,\Tr(\gamma_a + \gamma_b) = 0; \\
(-1)^{\frac{m(p-1)}{2}} p^m, & \text{if } \Tr(\gamma_a^{p^{k_1}+1} + \gamma_b^{p^{k_2}+1}) \neq 0 , \Tr(\gamma_a + \gamma_b) \neq 0.
\end{cases}
\]
2) If $\frac{m_p}{2} + c \neq 0$, then
\[
\Omega_3 =
\begin{cases}
(-1)^{\frac{m(p-1)}{2}}(1-p)p^m, &
  \begin{aligned}
    &\text{if } \Tr(\gamma_a^{p^{k_1}+1}+\gamma_b^{p^{k_2}+1})=0 ,\Tr(\gamma_a+\gamma_b)=0, \\
    &\quad\text{or } \Tr(\gamma_a^{p^{k_1}+1}+\gamma_b^{p^{k_2}+1})\neq 0, \Tr(\gamma_a+\gamma_b)=0 \\
    &\quad\text{and } -( \frac{m_p}{2}+c )\Tr(\gamma_a^{p^{k_1}+1}+\gamma_b^{p^{k_2}+1}) \in \rm NSQ, \\
    &\quad\text{or } \Tr(\gamma_a^{p^{k_1}+1}+\gamma_b^{p^{k_2}+1})\neq 0, \Tr(\gamma_a+\gamma_b)\neq 0 \text{ and } \Delta \in \rm NSQ;
  \end{aligned} \\[2ex]
(-1)^{\frac{m(p-1)}{2}}p^m, &
  \begin{aligned}
    &\text{if } \Tr(\gamma_a^{p^{k_1}+1}+\gamma_b^{p^{k_2}+1})=0, \Tr(\gamma_a+\gamma_b)\neq 0, \\
    &\quad\text{or } \Tr(\gamma_a^{p^{k_1}+1}+\gamma_b^{p^{k_2}+1})\neq 0, \Tr(\gamma_a+\gamma_b)\neq 0 \text{ and } \Delta=0;
  \end{aligned} \\[2ex]
(-1)^{\frac{m(p-1)}{2}}(p+1)p^m, &
  \begin{aligned}
    &\text{if } \Tr(\gamma_a^{p^{k_1}+1}+\gamma_b^{p^{k_2}+1})\neq 0, \Tr(\gamma_a+\gamma_b)=0 \\
    &\quad\text{and }- ( \frac{m_p}{2}+c )\Tr(\gamma_a^{p^{k_1}+1}+\gamma_b^{p^{k_2}+1}) \in \rm SQ, \\
    &\quad\text{or } \Tr(\gamma_a^{p^{k_1}+1}+\gamma_b^{p^{k_2}+1})\neq 0, \Tr(\gamma_a+\gamma_b)\neq 0 \text{ and } \Delta \in \rm SQ,
  \end{aligned}
\end{cases}
\]
where $\Delta = (\Tr(\gamma_a + \gamma_b))^2 - 4( \frac{m_p}{2} + c )\Tr(\gamma_a^{p^{k_1}+1}+\gamma_b^{p^{k_2}+1})$.

\end{lem}
\begin{proof}
By the assumption $2v_2(m) = v_2(u) + v_2(v)$, we have $v_2(m) = v_2(u) = v_2(v)$, which implies that $\frac{m}{u}$ and $\frac{m}{v}$ are odd. Note that $f_{k_1}(x) = z_1^{p^{k_1}}x^{p^{2{k_1}}} + z_1x$ and $f_{k_2}(x) = z_1^{p^{k_2}}x^{p^{2{k_2}}} + z_1x$ are permutations on $\mathbb{F}_{q}$ for any $z_1 \in \mathbb{F}_p^*$.
Accordingly, $\gamma_a$ and $\gamma_b$ are the unique solutions of $x^{p^{2{k_1}}} + x = -a^{p^{k_1}}$ and $x^{p^{2{k_2}}} + x = -b^{p^{k_2}}$, respectively.
Similar to the discussion in the proof of \cite[Lemma 10]{HQLT}, it is clear that
$z_1^{-1}( z_2\gamma_a - \frac{1}{2}z_1 )$ and  $z_1^{-1}( z_2\gamma_b - \frac{1}{2}z_1 )$ are the solutions of $f_{k_1}(x) = -(z_1 + z_2a)^{p^{k_1}}$ and $f_{k_2}(x) = -(z_1 + z_2b)^{p^{k_2}}$, respectively.

By Lemma \ref{lemma7}-1), we have
\begin{align*}
\Omega_3 &= \sum_{z_1 \in \mathbb{F}_p^*} \xi_p^{-z_1 c} \sum_{z_2 \in \mathbb{F}_p^*} S_{k_1}(z_1, z_1 + z_2 a)S_{k_2}(z_1, z_1 + z_2 b) \\
&= (-1)^{\frac{m(p-1)}{2}} p^m \sum_{z_1, z_2 \in \mathbb{F}_p^*}  \xi_p^{-z_1 c}\chi \big( -z_1 ( z_1^{-1} ( z_2 \gamma_a - \frac{1}{2} z_1 ) )^{p^{k_1}+1} \big)\chi \big( -z_1 ( z_1^{-1} ( z_2 \gamma_b - \frac{1}{2} z_1 ) )^{p^{k_2}+1} \big).
\end{align*}

Meanwhile, a detailed calculation gives $\Tr ((z_2 \gamma_a - \frac{1}{2} z_1)^{p^{k_1}+1})=\Tr(z^2_2\gamma_a^{p^{k_1}+1}+\frac{z^2_1}{4}-\frac{1}{2}z_1z_2
\gamma_a^{p^{k_1}}-\frac{1}{2}z_1z_2\gamma_a) = z_2^2 \Tr(\gamma_a^{p^{k_1}+1}) - z_1 z_2 \operatorname{Tr}(\gamma_a) + \frac{m_p z_1^2}{4},$
and $\Tr((z_2 \gamma_b - \frac{1}{2} z_1)^{p^{k_2}+1}) = z_2^2 \operatorname{Tr}(\gamma_b^{p^{k_2}+1}) - z_1 z_2 \Tr(\gamma_b) + \frac{m_p z_1^2}{4}$, which indicates

\begin{equation}\label{W4.6}
\Omega_3 = (-1)^{\frac{m(p-1)}{2}} p^m \sum_{z_1 \in \mathbb{F}_p^*} \xi_p^{- z_1 c} \sum_{z_2 \in \mathbb{F}_p^*} \xi_p^{-\frac{z_2^2}{z_1} \Tr ( \gamma_a^{p^{k_1}+1} + \gamma_b^{p^{k_2}+1} ) + z_2 \Tr (\gamma_a + \gamma_b) - z_1 \frac{m_p}{2}}.
\end{equation}

Now we consider the following six cases for $\Omega_3 $.

If $\Tr(\gamma_a^{p^{k_1}+1} + \gamma_b^{p^{k_2}+1}) = 0$ and $\Tr(\gamma_a + \gamma_b) = 0$, then
\begin{align}\label{W4.7}
\Omega_3 &= (-1)^{\frac{m(p-1)}{2}} p^m (p-1) \sum_{z_1 \in \mathbb{F}_p^*} \xi_p^{-z_1  ( \frac{m_p}{2} + c )} \nonumber\\
&=
\begin{cases}
(-1)^{\frac{m(p-1)}{2}} p^m (p-1)^2, & \text{if } \frac{m_p}{2} + c = 0; \\
-(-1)^{\frac{m(p-1)}{2}} p^m (p-1), & \text{if } \frac{m_p}{2} + c \neq 0 .
\end{cases}
\end{align}

If $\Tr(\gamma_a^{p^{k_1}+1} + \gamma_b^{p^{k_2}+1}) = 0$ and $\Tr(\gamma_a + \gamma_b) \neq 0$, then
\begin{align}\label{W4.8}
\Omega_3 &= (-1)^{\frac{m(p-1)}{2}} p^m \sum_{z_1\in \mathbb{F}_p^*}\sum_{z_2 \in \mathbb{F}_p^*} \xi_p^{ z_2 \Tr (\gamma_a + \gamma_b)-z_1  ( \frac{m_p}{2} + c )} \nonumber\\
&=
\begin{cases}
-(-1)^{\frac{m(p-1)}{2}} p^m (p-1), & \text{if } \frac{m_p}{2} + c = 0; \\
(-1)^{\frac{m(p-1)}{2}} p^m, & \text{if } \frac{m_p}{2} + c \neq 0.
\end{cases}
\end{align}

If $\Tr(\gamma_a^{p^{k_1}+1} + \gamma_b^{p^{k_2}+1}) \neq 0$, $\Tr(\gamma_a + \gamma_b) = 0$, and $\frac{m_p}{2} + c \neq 0$, then from Lemma \ref{lemma3} and the fact that $G_1^2(\eta_1) =\eta_1(-1)p$, we obtain
\begin{align}\label{W4.9}
\Omega_3 &= (-1)^{\frac{m(p-1)}{2}} p^m \sum_{z_1 \in \mathbb{F}_p^*} \xi_{p}^{-z_1 c} \sum_{z_2 \in \mathbb{F}_p^*} \xi_{p}^{-\frac{z_2^2}{z_1}\Tr ( \gamma_a^{p^{k_1}+1} + \gamma_b^{p^{k_2}+1} ) -z_1{\frac{m_p}{2}}}  \nonumber\\
&= (-1)^{\frac{m(p-1)}{2}} p^m \eta_1  (( c + \frac{m_p}{2} ) \Tr ( \gamma_a^{p^{k_1}+1} + \gamma_b^{p^{k_2}+1} )) G_1^2(\eta_1) + (-1)^{\frac{m(p-1)}{2}} p^m \nonumber\\
&= (-1)^{\frac{m(p-1)}{2}} p^{m+1} \eta_1  (-( c + \frac{m_p}{2} ) \Tr ( \gamma_a^{p^{k_1}+1} + \gamma_b^{p^{k_2}+1} )) + (-1)^{\frac{m(p-1)}{2}} p^m.
\end{align}

If $\Tr(\gamma_a^{p^{k_1}+1} + \gamma_b^{p^{k_2}+1}) \neq 0$, $\Tr(\gamma_a + \gamma_b) = 0$, and $\frac{m_p}{2} + c = 0$, then
\begin{equation}\label{W4.10}
\Omega_3 = -(-1)^{\frac{m(p-1)}{2}} p^m (p-1).
\end{equation}


If $\Tr(\gamma_a^{p^{k_1}+1} + \gamma_b^{p^{k_2}+1}) \neq 0$, $\Tr(\gamma_a + \gamma_b) \neq 0$, and $\frac{m_p}{2} + c \neq 0$, then $\Omega_3$ is equal to
\begin{align}\label{W4.11}
&(-1)^{\frac{m(p-1)}{2}} p^m \sum_{z_1 \in \mathbb{F}_p^*} \xi_p^{- z_1 c} \sum_{z_2 \in \mathbb{F}_p^*} \xi_p^{-\frac{z_2^2}{z_1} \Tr ( \gamma_a^{p^{k_1}+1} + \gamma_b^{p^{k_2}+1} ) + z_2 \Tr (\gamma_a + \gamma_b) - z_1 \frac{m_p}{2}} \nonumber\\
=&(-1)^{\frac{m(p-1)}{2}} p^m (\eta_1(- \Tr ( \gamma_a^{p^{k_1}+1} + \gamma_b^{p^{k_2}+1} ))G_1(\eta_1)\sum_{z_1 \in \mathbb{F}_p^*}\xi_p^{\left( \frac{(\Tr (\gamma_a + \gamma_b))^2}{4\Tr ( \gamma_a^{p^{k_1}+1} + \gamma_b^{p^{k_2}+1} )}-\frac{m_p}{2} - c\right)z_1}\eta_1(z_1)+1)\nonumber\\
=&
\begin{cases}
(-1)^{\frac{m(p-1)}{2}} p^m, & \text{if } \Delta = 0; \\
(-1)^{\frac{m(p-1)}{2}} p^m (p \cdot \eta_1(\Delta) + 1), & \text{if } \Delta \neq 0,
\end{cases}
\end{align}
where $\Delta = (\Tr (\gamma_a + \gamma_b))^2 - 4 ( \frac{m_p}{2} + c ) \Tr ( \gamma_a^{p^{k_1}+1} + \gamma_b^{p^{k_2}+1} )$.

If $\Tr(\gamma_a^{p^{k_1}+1} + \gamma_b^{p^{k_2}+1}) \neq 0$, $\Tr(\gamma_a + \gamma_b) \neq 0$, and $\frac{m_p}{2} + c = 0$, then
\begin{align}\label{W4.12}
\Omega_3 &= (-1)^{\frac{m(p-1)}{2}} p^m \sum_{z_1 \in \mathbb{F}_p^*} \sum_{z_2 \in \mathbb{F}_p^*} \xi_p^{-\frac{z_2^2}{z_1} \Tr ( \gamma_a^{p^{k_1}+1} + \gamma_b^{p^{k_2}+1} ) + z_2 \Tr (\gamma_a + \gamma_b)} = (-1)^{\frac{m(p-1)}{2}} p^m.
\end{align}

With the discussion above, the results follow. This completes the proof.
\end{proof}

\begin{lem}\label{lemma14}
Let $(a,b)\in\mathbb F_q^2\setminus\{(0,0)\}$. Let $\gamma_a$ and $\gamma_b$ be solutions of the equations  $x^{p^{2{k_1}}} + x = -a^{p^{k_1}}$ and $x^{p^{2{k_2}}} + x = -b^{p^{k_2}}$, respectively. Assume that $2v_2(m) = v_2(u) + v_2(v)+1$.\\
1) If $\frac{m_p}{2} + c = 0$, then
\[
\Omega_3 =
\begin{cases}
(-1)^{\frac{m(p-1)}{4}} (p-1)^2 p^m, & \text{if } \Tr(\gamma_a^{p^{k_1}+1} + \gamma_b^{p^{k_2}+1}) = 0 \text{ and } \Tr(\gamma_a + \gamma_b) = 0; \\
-(-1)^{\frac{m(p-1)}{4}} (p-1)p^m, & \text{if } \Tr(\gamma_a^{p^{k_1}+1} + \gamma_b^{p^{k_2}+1}) = 0 \text{ and } \Tr(\gamma_a + \gamma_b) \neq 0, \\
&\quad \text{or } \Tr(\gamma_a^{p^{k_1}+1} + \gamma_b^{p^{k_2}+1}) \neq 0 \text{ and } \Tr(\gamma_a + \gamma_b) = 0; \\
(-1)^{\frac{m(p-1)}{4}} p^m, & \text{if } \Tr(\gamma_a^{p^{k_1}+1} + \gamma_b^{p^{k_2}+1}) \neq 0 \text{ and } \Tr(\gamma_a + \gamma_b) \neq 0.
\end{cases}
\]
2) If $\frac{m_p}{2} + c \neq 0$, then
\[
\Omega_3 =
\begin{cases}
(-1)^{\frac{m(p-1)}{4}}(1-p)p^m, &
  \begin{aligned}
    &\text{if } \Tr(\gamma_a^{p^{k_1}+1}+\gamma_b^{p^{k_2}+1})=0 \text{ and } \Tr(\gamma_a+\gamma_b)=0, \\
    &\quad\text{or } \Tr(\gamma_a^{p^{k_1}+1}+\gamma_b^{p^{k_2}+1})\neq 0, \Tr(\gamma_a+\gamma_b)=0 \\
    &\quad\text{and } -( \frac{m_p}{2}+c )\Tr(\gamma_a^{p^{k_1}+1}+\gamma_b^{p^{k_2}+1}) \in \rm NSQ, \\
    &\quad\text{or } \Tr(\gamma_a^{p^{k_1}+1}+\gamma_b^{p^{k_2}+1})\neq 0, \Tr(\gamma_a+\gamma_b)\neq 0 \text{ and } \Delta \in \rm NSQ;
  \end{aligned} \\[2ex]
(-1)^{\frac{m(p-1)}{4}}p^m, &
  \begin{aligned}
    &\text{if } \Tr(\gamma_a^{p^{k_1}+1}+\gamma_b^{p^{k_2}+1})=0 \text{ and } \Tr(\gamma_a+\gamma_b)\neq 0, \\
    &\quad\text{or } \Tr(\gamma_a^{p^{k_1}+1}+\gamma_b^{p^{k_2}+1})\neq 0, \Tr(\gamma_a+\gamma_b)\neq 0 \text{ and } \Delta=0;
  \end{aligned} \\[2ex]
(-1)^{\frac{m(p-1)}{4}}(p+1)p^m, &
  \begin{aligned}
    &\text{if } \Tr(\gamma_a^{p^{k_1}+1}+\gamma_b^{p^{k_2}+1})\neq 0, \Tr(\gamma_a+\gamma_b)=0 \\
    &\quad\text{and } -( \frac{m_p}{2}+c )\Tr(\gamma_a^{p^{k_1}+1}+\gamma_b^{p^{k_2}+1}) \in \rm SQ, \\
    &\quad\text{or } \Tr(\gamma_a^{p^{k_1}+1}+\gamma_b^{p^{k_2}+1})\neq 0, \Tr(\gamma_a+\gamma_b)\neq 0 \text{ and } \Delta \in \rm SQ,
  \end{aligned}
\end{cases}
\]
where $\Delta = (\Tr(\gamma_a + \gamma_b))^2 - 4( \frac{m_p}{2} + c )\Tr(\gamma_a^{p^{k_1}+1}+\gamma_b^{p^{k_2}+1})$.

\end{lem}

\begin{proof}
Due to $2v_2(m) = v_2(u) + v_2(v)+1$, $m$ is even and there are two cases, which are given in Eq. \eqref{W2}.  Here we only prove the case $v_2(m) = v_2(u)$ and $v_2(m) = v_2(v)+1$. This implies that $\frac{m}{u}$ and $\frac{m}{2v}$ are odd. Note that $f_{k_1}(x) = z_1^{p^{k_1}}x^{p^{2{k_1}}} + z_1x$ and $f_{k_2}(x) = z_1^{p^{k_2}}x^{p^{2{k_2}}} + z_1x$
are permutations over $\mathbb{F}_{q}$ for any $z_1 \in \mathbb{F}_p^*$.
Similar to the discussion in the proof of Lemma \ref{lemma13}, it is clear that
$z_1^{-1}( z_2\gamma_a - \frac{1}{2}z_1 )$ and
$z_1^{-1}( z_2\gamma_b - \frac{1}{2}z_1 )$ are the solutions of
$f_{k_1}(x) = -(z_1 + z_2a)^{p^{k_1}}$ and $f_{k_2}(x) = -(z_1 + z_2b)^{p^{k_2}}$, respectively.

By Lemma \ref{lemma7}, we have
\begin{align*}
\Omega_3 &= \sum_{z_1 \in \mathbb{F}_p^*} \xi_p^{-z_1 c} \sum_{z_2 \in \mathbb{F}_p^*} S_{k_1}(z_1, z_1 + z_2 a) S_{k_2}(z_1, z_1 + z_2 b) \\
&= \sum_{z_1,z_2 \in \mathbb{F}_p^*} \xi_p^{-z_1 c} ( (-1)^{m-1} \sqrt{(-1)}^{\frac{(p-1)^2 3 m}{4}} \, p^{\frac{m}{2}} \, \eta(z_1) \chi (- z_1 ( z_1^{-1} ( z_2 \gamma_a - \frac{1}{2} z_1 ) )^{p^{k_1}+1} ) ) \\
&\quad\quad\quad\quad\quad ( (-1)^{\frac{m}{2v}} \, p^{\frac{m}{2}} \chi ( -z_1 ( z_1^{-1} ( z_2 \gamma_b - \frac{1}{2} z_1 ) )^{p^{k_2}+1} ) ) \\
&= (-1)^{\frac{m(p-1)}{4}} p^m \sum_{z_1, z_2 \in \mathbb{F}_p^*}  \xi_p^{-z_1 c} \chi \big( -z_1 ( z_1^{-1} ( z_2 \gamma_a - \frac{1}{2} z_1 ) )^{p^{k_1}+1} \big) \chi \big( -z_1 ( z_1^{-1} ( z_2 \gamma_b - \frac{1}{2} z_1 ) )^{p^{k_2}+1} \big).
\end{align*}

Similar to the discussion in Eqs. \eqref{W4.6}-\eqref{W4.12}, the results are obtained.
\end{proof}

\begin{lem}\label{lemma15}
Let $(a,b)\in\mathbb F_q^2\setminus\{(0,0)\}$ and assume that $2v_2(m)>v_2(u)+v_2(v)+1$. If either $x^{p^{2k_1}}+x=-a^{p^{k_1}}$ or $x^{p^{2k_2}}+x=-b^{p^{k_2}}$ has no solution in $\mathbb F_q$, then $\Omega_3=0$. Otherwise, let $\gamma_a$ and $\gamma_b$ be solutions of the two equations, respectively.\\
1) If $\frac{m_p}{2} + c = 0$, then
\[
\Omega_3 =
\begin{cases}
(p-1)^2 p^{m+\varepsilon_u u+\varepsilon_v v}, & \text{if } \, \Tr( \gamma_a^{p^{k_1}+1} + \gamma_b^{p^{k_2}+1}) = 0 \, \text{and } \, \Tr( \gamma_a + \gamma_b ) = 0; \\
-(p-1)p^{m+\varepsilon_u u+\varepsilon_v v}, & \text{if } \, \Tr(\gamma_a^{p^{k_1}+1} + \gamma_b^{p^{k_2}+1} ) = 0 \, \text{and } \, \Tr( \gamma_a + \gamma_b ) \neq 0, \\
& \quad\text{or } \, \Tr( \gamma_a^{p^{k_1}+1} + \gamma_b^{p^{k_2}+1} ) \neq 0 \, \text{and } \, \Tr( \gamma_a + \gamma_b ) = 0; \\
p^{m+\varepsilon_u u+\varepsilon_v v}, & \text{if } \, \Tr( \gamma_a^{p^{k_1}+1} + \gamma_b^{p^{k_2}+1} ) \neq 0 \, \text{and } \, \Tr( \gamma_a + \gamma_b ) \neq 0; \\
0, & \text{if } \, x^{p^{2k_1}} + x = -a^{p^{k_1}} \, \text{or }\,x^{p^{2k_2}} + x = -b^{p^{k_2}} \, \text{has no solution in } \mathbb{F}_{q}.
\end{cases}
\]
2) If $\frac{m_p}{2} + c \neq 0$, then
\[
\Omega_3 =
\begin{cases}
(1-p)p^{m+\varepsilon_u u+\varepsilon_v v}, &
\begin{aligned}
& \text{if } \Tr(\gamma_a^{p^{k_1}+1} + \gamma_b^{p^{k_2}+1} ) = 0 \text{ and } \Tr(\gamma_a+\gamma_b) = 0, \\
& \quad\text{or } \Tr(\gamma_a^{p^{k_1}+1} + \gamma_b^{p^{k_2}+1} ) \neq 0, \Tr(\gamma_a+\gamma_b) = 0 \text{ and } \\
& \quad- (\frac{m_p}{2}+c)\Tr(\gamma_a^{p^{k_1}+1}+\gamma_b^{p^{k_2}+1}) \in \mathrm{\rm NSQ}, \\
& \quad\text{or } \Tr(\gamma_a^{p^{k_1}+1} + \gamma_b^{p^{k_2}+1} ) \neq 0, \Tr(\gamma_a+\gamma_b) \neq 0 \text{ and } \Delta \in \mathrm{\rm NSQ};
\end{aligned} \\[2em]
p^{m+\varepsilon_u u+\varepsilon_v v}, &
\begin{aligned}
& \text{if } \Tr(\gamma_a^{p^{k_1}+1} + \gamma_b^{p^{k_2}+1} ) = 0 \text{ and } \Tr(\gamma_a+\gamma_b) \neq 0, \\
& \quad\text{or } \Tr(\gamma_a^{p^{k_1}+1} + \gamma_b^{p^{k_2}+1} ) \neq 0, \Tr(\gamma_a+\gamma_b) \neq 0 \text{ and } \Delta = 0;
\end{aligned} \\[2em]
(p+1)p^{m+\varepsilon_u u+\varepsilon_v v}, &
\begin{aligned}
& \text{if } \Tr(\gamma_a^{p^{k_1}+1} + \gamma_b^{p^{k_2}+1} ) \neq 0, \Tr(\gamma_a+\gamma_b) = 0 \text{ and } \\
& \quad-(\frac{m_p}{2}+c)\Tr(\gamma_a^{p^{k_1}+1} + \gamma_b^{p^{k_2}+1} ) \in \mathrm{SQ}, \\
& \quad\text{or } \Tr(\gamma_a^{p^{k_1}+1} + \gamma_b^{p^{k_2}+1} ) \neq 0, \Tr(\gamma_a+\gamma_b) \neq 0 \text{ and } \Delta \in \mathrm{SQ};
\end{aligned} \\[2em]
0, & \text{if } \, x^{p^{2k_1}} + x = -a^{p^{k_1}} \, \text{or }\, x^{p^{2k_2}} + x = -b^{p^{k_2}} \, \text{has no solution in } \mathbb{F}_{q},
\end{cases}
\]
where $\Delta = (\Tr(\gamma_a + \gamma_b))^2 - 4( \frac{m_p}{2} + c )\Tr(\gamma_a^{p^{k_1}+1}+\gamma_b^{p^{k_2}+1})$.

\begin{proof}
Due to $2v_2(m) > v_2(u) + v_2(v) + 1$, $m$ is even and there are six cases, which are given in Eq. \eqref{W1}.  Here we only prove the case $v_2(m) > v_2(u) + 1$ and $v_2(m) > v_2(v) + 1$. This shows that $\varepsilon_u = 1$ and $\varepsilon_v = 1$. Note that $f_{k_1}(x) = z_1^{p^{k_1}}x^{p^{2{k_1}}} + z_1x$ and $f_{k_2}(x) = z_1^{p^{k_2}}x^{p^{2{k_2}}} + z_1x$
are not permutations over $\mathbb{F}_{q}$ for any $z_1 \in \mathbb{F}_p^*$. In the following, we consider two cases. First, we suppose that $ z_1^{p^{k_1}} x^{p^{2{k_1}}} + z_1 x = -(z_1 + z_2 a)^{p^{k_1}}$  has no solution in $\mathbb{F}_{q}$ or  $z_1^{p^{k_2}} x^{p^{2{k_2}}} + z_1 x= -(z_1 + z_2 b)^{p^{k_2}}$ has no solution in $\mathbb{F}_{q}$. Then, by Lemma \ref{lemma8}, we have $\Omega_3 = 0$. Second, we assume that $\gamma_a$ and $\gamma_b$ are the solutions of the equations $ x^{p^{2{k_1}}} + x = -a^{p^{k_1}}$ and $ x^{p^{2{k_2}}} + x = -b^{p^{k_2}}$, respectively. Similar to the discussion in the proof of Lemma \ref{lemma13}, by Lemma \ref{lemma8}, we have
\begin{align*}
\Omega_3 &= \sum_{z_1 \in \mathbb{F}_p^*} \xi_p^{-z_1 c} \sum_{z_2 \in \mathbb{F}_p^*} S_{k_1}(z_1, z_1 + z_2 a) S_{k_2}(z_1, z_1 + z_2 b) \\
&= \sum_{z_1,z_2 \in \mathbb{F}_p^*} \xi_p^{-z_1 c} ( -(-1)^{\frac{m}{2u}} p^{\frac{m}{2} + u} \chi (- z_1 ( z_1^{-1} ( z_2 \gamma_a - \frac{1}{2} z_1 ) )^{p^{k_1}+1} ) ) \\
&\quad\quad \quad\quad\quad \quad \quad \quad  \cdot ( -(-1)^{\frac{m}{2v}} p^{\frac{m}{2} + v} \chi ( -z_1 ( z_1^{-1} ( z_2 \gamma_b - \frac{1}{2} z_1 ) )^{p^{k_2}+1} ) ) \\
&= p^{m+u+v} \sum_{z_1, z_2 \in \mathbb{F}_p^*}  \xi_p^{-z_1 c} \chi \big( -z_1 ( z_1^{-1} ( z_2 \gamma_a - \frac{1}{2} z_1 ) )^{p^{k_1}+1} \big) \chi \big( -z_1 ( z_1^{-1} ( z_2 \gamma_b - \frac{1}{2} z_1 ) )^{p^{k_2}+1} \big).
\end{align*}

Following a similar discussion to that in Eqs. \eqref{W4.6}-\eqref{W4.12}, the results are obtained.
\end{proof}
\end{lem}

Furthermore, to determine the weight distributions of the codes, we also need the following lemmas.
\begin{lem}\label{lemma16}
Let
\[
M= |\{(x, y) \in \mathbb{F}_q^2 \setminus \{(0, 0)\} : \Tr(x^{p^{k_1}+1} + y^{p^{k_2}+1}) = 0, \Tr(x + y) \neq 0\}|.
\]
1) If ${m_p}=0$, then
\[
M = p^{2m-2}(p-1).
\]
2) If ${m_p} \neq 0$, then
\[
M =
\begin{cases}
(p^{2m-2} + (-1)^{\frac{m(p-1)}{2}} p^{m-1})(p-1), & \text{if } 2v_2(m) = v_2(u) + v_2(v); \\
(p^{2m-2} + (-1)^{\frac{m(p-1)}{4}} p^{m-1})(p-1), & \text{if } 2v_2(m) = v_2(u) + v_2(v) + 1; \\
(p^{2m-2} + p^{m + \varepsilon_u u + \varepsilon_v v - 1})(p-1), & \text{if } 2v_2(m) > v_2(u) + v_2(v) + 1.
\end{cases}
\]
\begin{proof}
Let $g \in \mathbb{F}_p^*$. Define \[M_g=| \{(x, y) \in \mathbb{F}_q^2 \setminus \{(0, 0)\} : \Tr(x^{p^{k_1}+1} + y^{p^{k_2}+1}) = 0, \Tr(x + y) = g\}|.\]
Then we have
\begin{align}\label{W4.20}
M_g
&= \sum_{(x, y) \in \mathbb{F}_q^2 \setminus \{(0, 0)\}}  ( \frac{1}{p} \sum_{z_1 \in \mathbb{F}_p} \xi_p^{z_1 \Tr(x^{p^{k_1}+1} + y^{p^{k_2}+1})} ) ( \frac{1}{p} \sum_{z_2 \in \mathbb{F}_p} \xi_p^{z_2 (\Tr(x+y)-g)} )\nonumber\\
&= \frac{1}{p^2} \sum_{x, y \in \mathbb{F}_q} ( 1 + \sum_{z_1 \in \mathbb{F}_p^*} \xi_p^{z_1 \Tr(x^{p^{k_1}+1} + y^{p^{k_2}+1})} + \sum_{z_2 \in \mathbb{F}_p^*} \xi_p^{z_2 (\Tr(x+y)-g)} \nonumber\\
&\quad +\sum_{z_1 \in \mathbb{F}_p^*} \xi_p^{z_1 \Tr(x^{p^{k_1}+1} + y^{p^{k_2}+1})} \sum_{z_2 \in \mathbb{F}_p^*} \xi_p^{z_2 (\Tr(x+y)-g)} )- \frac{1}{p^2} \sum_{z_1,z_2 \in \mathbb{F}_p}\xi_p^{-z_2g}\nonumber\\
&= p^{2m-2} + \frac{1}{p^2} (M_1 + 0 + M_2),
\end{align}
where
\[M_1 = \sum_{x,y \in \mathbb{F}_q} \sum_{z_1 \in \mathbb{F}_p^*} \xi_{p}^{z_1 \Tr(x^{p^{k_1}+1} + y^{p^{k_2}+1})},\]
and
\[M_2 = \sum_{x,y \in \mathbb{F}_q} \sum_{z_2 \in \mathbb{F}_p^*} \xi_p^{z_2 (\Tr(x+y)-g)} \sum_{z_1 \in \mathbb{F}_p^*} \xi_p^{z_1 \Tr(x^{p^{k_1}+1} + y^{p^{k_2}+1})}.\]

Following the discussion in the proof of \cite[Proposition 11]{ZZWZ} and applying Lemma \ref{lemma2}, we have
\begin{align*}
M_1 &= \sum_{x,y \in \mathbb{F}_q} \sum_{z_1 \in \mathbb{F}_p^*} \xi_{p}^{z_1 \Tr(x^{p^{k_1}+1} + y^{p^{k_2}+1})} \\
&= \sum_{z_1 \in \mathbb{F}_p^*} \sum_{x \in \mathbb{F}_q} \xi_{p}^{z_1 \Tr(x^{p^{k_1}+1})} \sum_{y \in \mathbb{F}_q} \xi_{p}^{z_1 \Tr(y^{p^{k_2}+1})} \\
&= \sum_{z_1 \in \mathbb{F}_p^*} \eta_1^{r_1 + r_2} (z_1) \sum_{x \in \mathbb{F}_q} \xi_{p}^{ \Tr(x^{p^{k_1}+1})} \sum_{y \in \mathbb{F}_q} \xi_{p}^{\Tr(y^{p^{k_2}+1})},
\end{align*}
where $r_1$ and $r_2$ are the ranks of the quadratic forms $\Tr(x^{p^{k_1}+1})$ and $\Tr(x^{p^{k_2}+1})$, respectively. By Lemma \ref{lemma4}, it is easy to see that $\eta_1^{r_1 + r_2} (z_1)=1$ for any $z \in \mathbb{F}_p^*$ since $r_1 + r_2$ is even. Hence, it follows from Lemma \ref{lemma5} that
\[
M_1 =
\begin{cases}
(-1)^{\frac{m(p-1)}{2}} p^m (p-1), & \text{if } 2v_2(m) = v_2(u) + v_2(v); \\
(-1)^{\frac{m(p-1)}{4}} p^m (p-1), & \text{if } 2v_2(m) = v_2(u) + v_2(v) + 1; \\
p^{m+\varepsilon_u u + \varepsilon_v v} (p-1), & \text{if } 2v_2(m) > v_2(u) + v_2(v) + 1.
\end{cases}
\]

By an argument analogous to that used in the computation of $\Omega_3$ and in the proofs of Lemmas \ref{lemma13}, \ref{lemma14}, and \ref{lemma15}, we obtain the following results.

When ${m_p}=0$ , we have
\begin{align*}
M_2 &= \sum_{x,y \in \mathbb{F}_q} \sum_{z_2 \in \mathbb{F}_p^*} \xi_p^{z_2 (\Tr(x+y)-g)} \sum_{z_1 \in \mathbb{F}_p^*} \xi_p^{z_1 \Tr(x^{p^{k_1}+1} + y^{p^{k_2}+1})} \\
&=
\begin{cases}
-(-1)^{\frac{m(p-1)}{2}} p^m (p-1), & \text{if } 2v_2(m) = v_2(u) + v_2(v); \\
-(-1)^{\frac{m(p-1)}{4}} p^m (p-1), & \text{if } 2v_2(m) = v_2(u) + v_2(v) + 1; \\
-p^{m+\varepsilon_u u + \varepsilon_v v} (p-1), & \text{if } 2v_2(m) > v_2(u) + v_2(v) + 1.
\end{cases}
\end{align*}

When ${m_p} \neq 0$ , we have
\begin{align*}
M_2 &= \sum_{x,y \in \mathbb{F}_q} \sum_{z_2 \in \mathbb{F}_p^*} \xi_p^{z_2 (\Tr(x+y)-g)} \sum_{z_1 \in \mathbb{F}_p^*} \xi_p^{z_1 \Tr(x^{p^{k_1}+1} + y^{p^{k_2}+1})} \\
&=
\begin{cases}
(-1)^{\frac{m(p-1)}{2}} p^m , & \text{if } 2v_2(m) = v_2(u) + v_2(v); \\
(-1)^{\frac{m(p-1)}{4}} p^m , & \text{if } 2v_2(m) = v_2(u) + v_2(v) + 1; \\
p^{m+\varepsilon_u u + \varepsilon_v v} , & \text{if } 2v_2(m) > v_2(u) + v_2(v) + 1.
\end{cases}
\end{align*}

Substituting $M_1$ and $M_2$ into Eq. \eqref{W4.20}, we obtain the value of $M_g$. Then
\[
M= \sum_{g \in \mathbb{F}_p^*}M_g= (p - 1)M_g.
\]
This completes the proof.
\end{proof}
\end{lem}

\begin{lem}\label{lemma17}
Let
\[
H= |\{(x, y) \in \mathbb{F}_q^2\setminus \{(0, 0)\}  : \Tr(x^{p^{k_1}+1} + y^{p^{k_2}+1}) = 0, \Tr(x + y) = 0\}|.
\]
1) If ${m_p}=0$, then
\[
H =
\begin{cases}
p^{2m-2} + (-1)^{\frac{m(p-1)}{2}} (p-1)p^{m-1}-1, & \text{if } 2v_2(m) = v_2(u) + v_2(v); \\
p^{2m-2} + (-1)^{\frac{m(p-1)}{4}} (p-1)p^{m-1}-1, & \text{if } 2v_2(m) = v_2(u) + v_2(v) + 1; \\
p^{2m-2} + (p-1)p^{m+\varepsilon_u u + \varepsilon_v v-1}-1, & \text{if } 2v_2(m) > v_2(u) + v_2(v) + 1.
\end{cases}
\]
2) If ${m_p} \neq 0$, then
\[
H = p^{2m-2} - 1.
\]

\begin{proof}
Similar to the proof of Lemma \ref{lemma16}, the results follow directly.
\end{proof}
\end{lem}

\begin{lem}\label{lemma18}
Let $a_1, a_2 \in \mathbb{F}_p^*$. Define
\[
T = |\{(x, y) \in \mathbb{F}_q^2 : \Tr(x^{p^{k_1}+1} + y^{p^{k_2}+1}) + \frac{a_1}{a_2} (\Tr(x + y))^2 = 0\}|.
\]
Then we have
\[
T =
\begin{cases}
p^{2m-1} + (-1)^{\frac{(p-1)m}{2}} (p-1)p^{m-1}\eta_1 ( 1 + 2\frac{a_1m_p}{a_2} ), & \text{if } 2v_2(m) = v_2(u) + v_2(v); \\
p^{2m-1} + (-1)^{\frac{(p-1)m}{4}} (p-1)p^{m-1}\eta_1 ( 1 + 2\frac{a_1m_p}{a_2} ), & \text{if } 2v_2(m) = v_2(u) + v_2(v) + 1; \\
p^{2m-1} + (p-1)p^{m+\varepsilon_u u + \varepsilon_v v-1}\eta_1 ( 1 + 2\frac{a_1m_p}{a_2} ), & \text{if } 2v_2(m) > v_2(u) + v_2(v) + 1.
\end{cases}
\]
\begin{proof}
Here we only give the proof for the case $2v_2(m) = v_2(u) + v_2(v)$, since the other cases can be proved similarly.
Similar to the proof of \cite[Lemma 16]{TXF}, we have
\begin{align}\label{W4.5}
T &= \frac{1}{p^2} \sum_{x,y \in \mathbb{F}_{q}} \sum_{\pi \in \mathbb{F}_p} ( \sum_{\omega \in \mathbb{F}_p} \chi_1(\omega(\pi - \Tr(x + y))) ) ( \sum_{z \in \mathbb{F}_p} \chi_1 ( z ( \Tr( x^{p^{k_1+1}} + y^{p^{k_2+1}} ) + \frac{a_1 \pi^2}{a_2} ) ) )\nonumber\\
&= \frac{1}{p^2} \sum_{x,y \in \mathbb{F}_{q}} \sum_{\pi \in \mathbb{F}_p} ( \sum_{\omega \in \mathbb{F}_p} \chi_1(\omega(\pi - \Tr(x + y))) ) + \nonumber\\
&\quad \frac{1}{p^2} \sum_{z \in \mathbb{F}_p^*} \sum_{\omega, \pi \in \mathbb{F}_p} \sum_{x,y \in \mathbb{F}_{q}} \chi_1 ( z\Tr( x^{p^{k_1+1}} + y^{p^{k_2+1}} ) + z \frac{a_1 \pi^2}{a_2} + \omega(\pi - \Tr(x + y)) )\nonumber\\
&= p^{2m-1} + T_1,
\end{align}
where
\begin{align*}
T_1 &= \frac{1}{p^2} \sum_{z \in \mathbb{F}_p^*} \sum_{\omega, \pi \in \mathbb{F}_p} \sum_{x, y \in \mathbb{F}_{q}} \chi_1 ( z\Tr ( x^{p^{k_1}+1} + y^{p^{k_2}+1} ) + z \frac{a_1 \pi^2}{a_2} + \omega (\pi - \Tr (x + y)) ) \\
&= \frac{1}{p^2} \sum_{z \in \mathbb{F}_p^*} \sum_{\omega, \pi \in \mathbb{F}_p} \chi_1 ( z \frac{a_1 \pi^2}{a_2} + \omega \pi ) \sum_{x, y \in \mathbb{F}_{q}} \chi ( z ( x^{p^{k_1}+1} + y^{p^{k_2}+1} ) - \omega (x + y) ) \\
&= \frac{1}{p^2} \sum_{z \in \mathbb{F}_p^*} \sum_{\omega, \pi \in \mathbb{F}_p} \chi_1 ( z \frac{a_1 \pi^2}{a_2} + \omega \pi ) \sum_{x \in \mathbb{F}_{q}} \chi ( zx^{p^{k_1}+1} - \omega x ) \sum_{y \in \mathbb{F}_{q}} \chi ( zy^{p^{k_2}+1} - \omega y ).
\end{align*}

By Lemma \ref{lemma7}, we have
\begin{align*}
T_1 &= \frac{1}{p^2} \sum_{z \in \mathbb{F}_p^*} \sum_{\omega, \pi \in \mathbb{F}_p} \chi_1 ( z \frac{a_1 \pi^2}{a_2} + \omega \pi ) \sum_{x \in \mathbb{F}_{q}} \chi ( zx^{p^{k_1}+1} - \omega x ) \sum_{y \in \mathbb{F}_{q}} \chi ( zy^{p^{k_2}+1} - \omega y ) \\
&= (-1)^{\frac{(p-1)m}{2}} p^{m-2} \sum_{z \in \mathbb{F}_p^*} \sum_{\omega, \pi \in \mathbb{F}_p} \chi_1 ( z \frac{a_1 \pi^2}{a_2} + \omega \pi - \frac{m_p \omega^2}{2z} ).
\end{align*}
This together with Lemma \ref{lemma3} and the fact $G_1^2(\eta_1) =\eta_1(-1)p$ gives
\begin{align*}
T_1 &= (-1)^{\frac{(p-1)m}{2}} p^{m-2} \sum_{z \in \mathbb{F}_p^*} \sum_{\omega \in \mathbb{F}_p} G_1(\eta_1) \eta_1 ( \frac{a_1 z}{a_2} ) \chi_1 ( -\frac{m_p \omega^2}{2z} - \frac{a_2 \omega^2}{4a_1 z} ) \\
&= (-1)^{\frac{(p-1)m}{2}} p^{m-2} \sum_{z \in \mathbb{F}_p^*} \sum_{\omega \in \mathbb{F}_p} G_1^2(\eta_1) \eta_1 ( \frac{a_1 z}{a_2} ) \eta_1 ( -\frac{m_p}{2z} - \frac{a_2}{4a_1 z} ) \\
&= (-1)^{\frac{(p-1)m}{2}} p^{m-2+1} \sum_{z \in \mathbb{F}_p^*} \eta_1 ( \frac{a_1 z}{a_2} ) \eta_1 ( \frac{m_p}{2z} + \frac{a_2}{4a_1 z} ) \\
&= (-1)^{\frac{(p-1)m}{2}} (p-1)p^{m-1} \eta_1 ( 1 + 2 \frac{a_1 m_p}{a_2} ).
\end{align*}
Substituting $T_1$ into Eq. \eqref{W4.5} yields the final result. It completes the proof.
\end{proof}
\end{lem}

\subsection{The proofs of Theorems \ref{thm1}, \ref{thm22} and \ref{thm33}}
Here we only prove the case $c \in \mathbb{F}_p^*$ in Theorem \ref{thm33}, which implies $\delta_c=0$. The remaining parts of Theorems \ref{thm1}, \ref{thm22} and \ref{thm33} can be proved  similarly.

When $2v_2(m) > v_2(u) + v_2(v) + 1$, then $m$ is even and there are six cases, which are given by Eq. \eqref{W1}. In the following, we only prove the case $v_2(m) > v_2(u) + 1$ and $v_2(m) > v_2(v) + 1$. The other cases can be proved similarly. This shows that $\varepsilon_u = 1$ and $\varepsilon_v = 1$. By Lemma \ref{lemma12}, the length of the code $\C_D$ is
  \begin{align}
  n&=\left\{\begin{array}{ll}
p^{2m-1} + p^{m+ u+ v-1}(p-1), & {\rm if} \,\, \frac{m_p}{2} + c = 0;\\
p^{2m-1} - p^{m+ u+ v-1}, & {\rm if} \,\, \frac{m_p}{2} + c \neq 0.
\end{array}\right.\nonumber
  \end{align}

In what follows, we study the weight distribution of $\C_D$. Recalling that the Hamming weight of a codeword $c(a, b)$ in $\C_D$ is defined as
\begin{equation}\label{wt}
wt (c(a, b))  = n - N(a, b).
\end{equation}
From Lemma \ref{lemma15}, we obtain a complete description of the value of $\Omega_3$ in this case. Therefore, by substituting the results of $\Omega_3$ and Eqs. \eqref{W4.3}-\eqref{W4.4} into Eq. \eqref{W4.2}, the value of $N(a, b)$ is completely determined. Then by Eq. \eqref{wt}, we can compute $wt (c(a, b)) $ as follows.

When $(a,b) = (0,0)$, we have $wt (c(a, b)) =  0$. Next, we compute $wt (c(a, b))$ for $(a,b) \neq (0,0)$. Here we first focus on the case $\frac{m_p}{2} + c = 0$. By  Lemmas \ref{lemma12}-3) and \ref{lemma15}-1) together with Eq. \eqref{wt}, $wt (c(a, b))$ is equal to
\[
\begin{cases}
w_1 :=(p-1)p^{2m-2}, & \text{if } \, \Tr( \gamma_a^{p^{k_1}+1} + \gamma_b^{p^{k_2}+1}) = 0, \Tr( \gamma_a + \gamma_b ) = 0; \\
w_2 :=(p-1)(p^{2m-2} + p^{m+u+v-1}), & \text{if } \, \Tr(\gamma_a^{p^{k_1}+1} + \gamma_b^{p^{k_2}+1} ) = 0, \Tr( \gamma_a + \gamma_b ) \neq 0, \\
& \quad\text{or } \, \Tr( \gamma_a^{p^{k_1}+1} + \gamma_b^{p^{k_2}+1} ) \neq 0, \Tr( \gamma_a + \gamma_b ) = 0; \\
w_3 :=(p-1)p^{2m-2} + (p-2)p^{m+u+v-1}, & \text{if } \, \Tr( \gamma_a^{p^{k_1}+1} + \gamma_b^{p^{k_2}+1} ) \neq 0, \Tr( \gamma_a + \gamma_b ) \neq 0; \\
w_4 :=(p-1)(p^{2m-2} + (p-1)p^{m+u+v-2}), & \text{if } \, x^{p^{2k_1}} + x = -a^{p^{k_1}} \, \text{or }\, x^{p^{2k_2}} + x = -b^{p^{k_2}} \text{ has} \\
& \quad\text{no solution in } \mathbb{F}_{q}.
\end{cases}
\]
It can be readily verified that $w_2 > w_4 > w_3 > w_1$ and $w_1 > 0$. This means that the minimum distance $d$ of $\C_D$ is equal to $p^{2m-1} - p^{2m-2}$ and the dimension of code $\C_D$ is equal to $2m$. By Lemmas \ref{lemma9} and \ref{lemma17}, we have $A_{w_1} = p^{2m-2(u+v)-2} - 1$, and
\begin{align}\label{Awt}
A_{w_4} &= | \{ (a, b) \in \mathbb{F}_{q}^2 \setminus \{(0, 0)\} :
\begin{aligned}
& x^{p^{2k_1}} + x = -a^{p^{k_1}} \text{ and } x^{p^{2k_2}} + x = -b^{p^{k_2}} \text{has no solution in } \mathbb{F}_{q}
\end{aligned} \} | \nonumber \\
&\quad + | \{ (a, b) \in \mathbb{F}_{q}^2 \setminus \{(0, 0)\} :
\begin{aligned}
& x^{p^{2k_1}} + x = -a^{p^{k_1}} \text{ has solution in } \mathbb{F}_{q} \\
& \text{and } x^{p^{2k_2}} + x = -b^{p^{k_2}} \text{ has no solution in } \mathbb{F}_{q}
\end{aligned} \} | \nonumber \\
&\quad + | \{ (a, b) \in \mathbb{F}_{q}^2 \setminus \{(0, 0)\} :
\begin{aligned}
& x^{p^{2k_1}} + x = -a^{p^{k_1}} \text{ has no solution in } \mathbb{F}_{q} \\
& \text{and } x^{p^{2k_2}} + x = -b^{p^{k_2}} \text{ has solution in } \mathbb{F}_{q}
\end{aligned} \} | \nonumber \\
&= (p^m - p^{m-2u})(p^m - p^{m-2v}) + p^{m-2u}(p^m - p^{m-2v}) + (p^m - p^{m-2u})p^{m-2v} \nonumber \\
&= p^{2m} - p^{2m-2u-2v}.
\end{align}
Since $(0,0)\notin D$, the minimum distance of $\C^\perp_{D}$ is bigger than one, i.e., $A^\perp_1=0$. From the first two Pless power moments (see \cite{Huff}, p.256), we have $ {A}_{{w}_{2}} = (p-1)(2 p^{2m-2(u+v)-2} + p^{m-(u+v)-1})$, and $ {A}_{{w}_{3}} = (p-1)^2p^{2m-2(u+v)-2}- (p-1)p^{m-(u+v)-1}$.

Next, for the case $\frac{m_p}{2} + c \neq 0$, we need to consider whether $m_p=0$ and whether $\eta_1 ( \frac{2c}{m_p + 2c} )=1$ or $-1$ for further calculation. We only prove the case $\frac{m_p}{2} + c \neq 0$, ${m_p} \neq 0$ and $\eta_1 ( \frac{2c}{m_p + 2c} )=1$, which implies $\sigma_c=1$, since the remaining cases can be proved  similarly.

By Lemmas \ref{lemma12}-3) and \ref{lemma15}-2) together with Eq. \eqref{wt}, $wt (c(a, b))$ is equal to
\[
\begin{cases}
w_1 := (p-1)p^{2m-2}, &
\begin{aligned}
& \text{if } \Tr(\gamma_a^{p^{k_1}+1} + \gamma_b^{p^{k_2}+1} ) = 0 , \Tr(\gamma_a+\gamma_b) = 0, \\
& \quad\text{or } \Tr(\gamma_a^{p^{k_1}+1} + \gamma_b^{p^{k_2}+1} ) \neq 0, \Tr(\gamma_a+\gamma_b) = 0,\\
& \quad- (\frac{m_p}{2}+c)\Tr(\gamma_a^{p^{k_1}+1}+\gamma_b^{p^{k_2}+1}) \in \mathrm{\rm NSQ}, \\
& \quad\text{or } \Tr(\gamma_a^{p^{k_1}+1} + \gamma_b^{p^{k_2}+1} ) \neq 0, \Tr(\gamma_a+\gamma_b) \neq 0, \Delta \in \mathrm{\rm NSQ};
\end{aligned} \\[2em]
w_2 := (p-1)p^{2m-2} - p^{m+u+v-1}, &
\begin{aligned}
& \text{if } \Tr(\gamma_a^{p^{k_1}+1} + \gamma_b^{p^{k_2}+1} ) = 0 \text{ and } \Tr(\gamma_a+\gamma_b) \neq 0, \\
& \quad\text{or } \Tr(\gamma_a^{p^{k_1}+1} + \gamma_b^{p^{k_2}+1} ) \neq 0, \Tr(\gamma_a+\gamma_b) \neq 0, \Delta = 0;
\end{aligned} \\[2em]
w_3 := (p-1)p^{2m-2}  - 2p^{m+u+v-1}, &
\begin{aligned}
& \text{if } \Tr(\gamma_a^{p^{k_1}+1} + \gamma_b^{p^{k_2}+1} ) \neq 0, \Tr(\gamma_a+\gamma_b) = 0,\\
& \quad-(\frac{m_p}{2}+c)\Tr(\gamma_a^{p^{k_1}+1} + \gamma_b^{p^{k_2}+1} ) \in \mathrm{SQ}, \\
& \quad\text{or } \Tr(\gamma_a^{p^{k_1}+1} + \gamma_b^{p^{k_2}+1} ) \neq 0, \Tr(\gamma_a+\gamma_b) \neq 0, \Delta \in \mathrm{SQ};
\end{aligned} \\[2em]
w_4 := (p-1)(p^{2m-2}- p^{m+u+v-2}), & \text{if } \, x^{p^{2k_1}} + x = -a^{p^{k_1}} \, \text{or }\,  x^{p^{2k_2}} + x = -b^{p^{k_2}} \text{ has no } \\
& \quad\text{ solution in } \mathbb{F}_{q}.
\end{cases}
\]
where $\Delta = (\Tr(\gamma_a+\gamma_b))^2 - 4(\frac{m_p}{2}+c)\Tr(\gamma_a^{p^{k_1}+1} + \gamma_b^{p^{k_2}+1})$.
Note that $w_1 > w_4 > w_2 > w_3$ and that $w_3 > 0$ provided that $m \geq  u +  v + 2$. Then the minimum distance $d$ of $\C_D$ is  $w_3$ and the dimension of code $\C_D$ is $2m$. By Eq. \eqref{Awt}, $A_{w_4} = p^{2m} - p^{2m-2u-2v}$. We now compute $ A_{w_2}$. Note that $\Delta = 0$ implies $\Tr(\gamma_a+\gamma_b)=0$, $\Tr(\gamma_a^{p^{k_1}+1} + \gamma_b^{p^{k_2}+1})=0$ or $\Tr(\gamma_a+\gamma_b)\neq 0$, $\Tr(\gamma_a^{p^{k_1}+1} + \gamma_b^{p^{k_2}+1})\neq 0$. By Lemmas \ref{lemma9}, \ref{lemma16}, \ref{lemma17} and \ref{lemma18}, we have $ A_{w_2}= (p-1)(p^{2m-2(u+v)-2} + p^{m-(u+v)-1})+p^{2m-2(u+v)-1}+(p-1)p^{m-(u+v)-1}-p^{2m-2(u+v)-2} =2(p-1)(p^{2m-2(u+v)-2} + p^{m-(u+v)-1})$. From Pless power moments (see \cite{Huff}, p.256), we have $A_{w_1} = \frac{1}{2}((p^2 - p + 2) p^{2m-2(u+v)-2}- (p-1) p^{m-(u+v)-1}) - 1$ and $A_{w_3} = \frac{1}{2}(p-1)((p-2)p^{2m-2(u+v)-2}-3p^{m-(u+v)-1})$. This completes the proof.

\section{Minimal linear codes} \label{s5}

In this section, we investigate the minimality of the codes $\C_D$ constructed in this paper and show that most of them are minimal.

Now, let us introduce some basic concepts and conclusions regarding minimal linear codes. Recall that a vector \( x \in \mathbb{F}_p^n \) covers a vector \( y \in \mathbb{F}_p^n \) if \( \operatorname{Suppt}(y) \subseteq \operatorname{Suppt}(x) \). A codeword \( x \) in \( C \) over \( \mathbb{F}_p \) is said to be minimal if \( x \) covers only the codeword \( a x \) for all \( a \in \mathbb{F}_p \), but no other codewords in \( C \). A linear code \( C \) over \( \mathbb{F}_p \) is said to be minimal if every codeword in \( C \) is minimal.

\begin{theorem}\label{thm4}
Let $\C_D$ be a linear code defined by \eqref{CD-1} and \eqref{D-2}.
\begin{itemize}
\item [1)] For code $\C_D$ given by Theorems \ref{thm1} and \ref{thm22}, $\C_D$ is minimal if $m \geq 3$.
\item [2)]  For code $\C_D$ given by Theorem \ref{thm33}-1), $\C_D$ is minimal if $m > \varepsilon_u u + \varepsilon_v v + 1$.
\item [3)]  For code $\C_D$ given by Theorem \ref{thm33}-2), $\C_D$ is minimal if $m \geq \varepsilon_u u + \varepsilon_v v + 2$ and $(m, p) \neq (\varepsilon_u u + \varepsilon_v v + 2, 3)$.
\end{itemize}
\begin{proof}
Here we only prove part 3) since other parts can be proved similarly. Let $w_{\min}$ and $w_{\max}$ be the minimum and maximum nonzero weights of $\C_D$, respectively. For the linear code $\C_D$ of Theorems \ref{thm33}-2), we have
\[
\frac{w_{\min}}{w_{\max}} = \frac{p^{2m-1} - p^{2m-2} - 2p^{m+\varepsilon_u u + \varepsilon_v v-1}}{p^{2m-1} - p^{2m-2}}.
\]
A direct calculation shows that $\frac{w_{\min}}{w_{\max}} > \frac{p-1}{p}$ holds if and only if $m \geq \varepsilon_u u + \varepsilon_v v + 2$ and $(m, p) \neq (\varepsilon_u u + \varepsilon_v v + 2, 3)$.
By Lemma \ref{lemma11}, it is shown that $\C_D$ is minimal. This completes the proof.
\end{proof}
\end{theorem}

\begin{remark}
Note that in Theorem \ref{thm4}, the codes $\C_D$ are minimal for sufficiently large $m$. Thus, most of our codes are minimal.
\end{remark}
\begin{remark} Some minimal codes are given as follows:
\begin{itemize}
\item [1)] The codes in Examples 1, 2 and 4 are minimal by Lemma \ref{lemma11}, consistent with our result in 1) of  Theorem~\ref{thm4}.
\item [2)]The code in Example 5 is minimal by Lemma \ref{lemma11}, consistent with our result in 2) of  Theorem~\ref{thm4}.
\item [3)]The code in Example 6 is minimal by Lemma \ref{lemma11}, consistent with our result in 3) of  Theorem~\ref{thm4}.
\end{itemize}
\end{remark}

\section{Concluding remarks} \label{s6}
This paper investigated the $p$-ary linear codes defined by \eqref{CD-1} and \eqref{D-2}. Several classes of three-weight or four-weight linear codes were derived. The parameters and weight distributions of the codes were completely determined through the theory of quadratic forms and Weil sums. In addition, it was shown that most of our linear codes are minimal. Moreover, some examples were provided via MAGMA computations, which are consistent with all our results.

\section*{Acknowledgements}
This work was supported by the National Natural Science Foundation of China (No. 12401688) and the Natural Science Foundation of Hubei Province of China (No. 2024AFB419).

\end{document}